\documentclass[review,nopreprintline]{elsarticle}
\usepackage{lineno}

\usepackage{amsmath}
\usepackage{subcaption}
\usepackage{mathrsfs}
\usepackage{algorithm}
\usepackage{algorithmic}
\usepackage{booktabs}
\usepackage{fontspec}
\usepackage{xeCJK}  

\usepackage{hyperref}
\journal{}

\usepackage[final]{changes} 
\definechangesauthor[name={Reviewer1}, color=blue]{R1}
\definechangesauthor[name={Reviewer2}, color=teal]{R2}
\definechangesauthor[name={Editor}, color=red]{E}
\definechangesauthor[name={Author}, color=gray]{A}

\begin{document}

\begin{frontmatter}

\title{Reduced-order modelling of parametrized unsteady Navier–Stokes equations and application to flow around cylinders with periodic changing boundary conditions}

\author{Shan Ding(丁杉)}

\author{Yongfu Tian(田永福)}

\author{Rui Yang(杨锐) \corref{mycorrespondingauthor}}
\cortext[mycorrespondingauthor]{Corresponding author}
\ead{ryang@tsinghua.edu.cn}

\address{School of Safety Science, Tsinghua University, Beijing, 100084, China}

\begin{abstract}
Computational fluid dynamics (CFD) simulations play an important role in engineering science and applications, however, it is not applicable for problems requiring a large number of repeated calculations. Accordingly, many reduced-order modelling techniques are developed to reduce computational costs, improve the efficiency, and have achieved significant progress. At present, most studies are focus on reconstructing the flow field throughout the parameter space of the snapshots on a fixed time window. However, the prediction problem has always been challenging, especially for unsteady flow. In this work, a reduced-order model (ROM) based on proper orthogonal decomposition (POD) and radial basis function (RBF) is presented and applied to the prediction problem of an unsteady flow with periodic changing boundary conditions. The method is validated by a numerical case of three-dimensional unsteady flow around cylinders with time-varying inlet velocity. This method is demonstrated to be quite accurate and efficient, reducing the CPU time by more than 99\% with an accuracy loss less than 5.2\% for predictions.
\end{abstract}

\begin{keyword}
\texttt Finite-volume Method \sep Reduced-order Modeling \sep Proper Orthogonal Decomposition \sep Unsteady Flow Field \sep Radial Basis Functions
\end{keyword}

\end{frontmatter}


\section{Introduction}

Although the numerical methods have made significant progress in solving Navier-Stokes equations, they still face great difficulties in practical applications. The computational cost will become unacceptably high when dealing with, for example, optimization and uncertainty analysis problems, which require a vast number of repeated calculations with different initial/boundary conditions. Similarly, \replaced[id=E]{solving equations is time-consuming and}{the time-consuming process of solving a system of equations} can be a significant challenge when predictions need to be obtained very quickly (e.g., predictions for emergency management).

Therefore, extensive researches have been conducted on reduced-order modelling techniques. The proper orthogonal decomposition (POD), also known as the Karhunen-Loève decomposition (KLD), the principal component analysis (PCA), and the singular value decomposition (SVD) is the most commonly used method to obtaining low-dimensional approximate descriptions of a high-dimensional system. Liang et al.\cite{LIANG2002527} discussed the equivalence between the three POD methods used in various disciplines. The main idea is to capture the dominant components from calculated solutions or experimental data. The principal components are also referred to as basis functions, proper orthogonal modes, basis vectors, empirical eigenfunctions, empirical basis functions or empirical orthogonal functions\cite{LIANG2002527}. 

In fluid mechanics, the POD method has two typical applications. One is to study the coherent structures of turbulent flows, and the other is to construct the low-dimensional model of the high-precision flow field. Cordier et al.\cite{2008two} introduced these two applications in detail and verified them with numerical examples. Berkooz et al.\cite{1994The} first led the POD method into the study of turbulent coherent structure. They projected the governing equations into the subspace constructed on POD modes, and transformed the infinite-dimensional hydrodynamic system into a low-dimensional model. Subsequently, Holmes et al.\cite{1996Turbulence} successfully identified and predicted the coherent structure of turbulent shear flow by this method. Since then, this method has been widely used in computational fluid dynamics (CFD). It has applied in many different engineering fields where it has been studied for air or water flows, such as heated planar jet\cite{TERASHIMA2018113}, circular jet mixing\cite{HE2018245}, vortex shedding around circular cylinder\cite{2017POD}, turbulent flows in urban street canyons\cite{FANG201496}, internal flow in turbine machinery\cite{cizmas2003proper}, multiphase porous media flows\cite{xiao2017non}. Trieu et al.\cite{trieu2022proper} analyzed the vortex structure behind the micro-vortex generator by POD, using the velocity and Liutex(the rotation strength) as inputs, respectively. Star et al.\cite{star2021pod} used POD to analyze the turbulent convective buoyant flow of sodium, reconstructing the dynamics \deleted[id=E]{process} and heat transfer \deleted[id=E]{process} accurately. The POD-Galerkin method is frequently used because of its versatility and adaptability. Barone et al.\cite{2008galerkin,2009Stable} analyzed the numerical instability of POD-Galerkin method and applied it to the reduced order modelling of linearized compressible flow. Stabile et al.\cite{2017POD,STABILE2018273} performed mathematical derivations of POD-Galerkin projection in detail, compared the two ROMs using continuity equation and pressure Poisson equation as control equations, and discussed how to ensure the numerical stability of the pressure solutions.

In addition to numerical instability, another disadvantage of the POD-Galerkin method is non-linearity inefficiency\cite{FRANCA1992209}. POD-Galerkin cannot solve the problems described in the form of discontinuous functions. Integral evaluation of the whole computational domain is required during reconstruction, which is computationally expensive\cite{2009A}. Consequently, some alternative approaches were developed, such as the Taylor series, Smolyak sparse grid collocation and radial basis function (RBF) interpolation. Xiao et al.\cite{2015Non} proposed a ROM based on the Taylor series method and a ROM  based on the Smolyak sparse grid method, and  compared them on two test cases. Radial basis function (RBF) interpolation is more efficient, convenient, and accurate for high-dimensional scattered data.

Qamar et al.\cite{QAMAR20091218} interpolated and extrapolated the POD mode coefficients to predict the steady supersonic flow-field of a new solution at different parameter values. Walton et al.\cite{2013Reduced} applied a second-level POD and established ROM for unsteady compressible inviscid flow over an oscillating ONERA M6 wing. They compared a variety of interpolation methods and proved the superiority of RBF. Xiao et al.\cite {XIAO201635} proposed the POD-RBF method and illustrated it by three simulations of fluid–structure interaction. Xiao et al.\cite{2016A} applied the POD-RBF method to obtain ROM of solid interacting with compressible fluid flows, and reconstructed the crack initiation and propagation during blasting. They discussed whether or not subtracting the mean field from the snapshots before applying POD will affect the reconstruction effect. \replaced[id=E]{To}{In order to} provide the flexibility to choose different numbers of POD modes and improve local accuracy of the complex flow, Xiao et al.\cite{XIAO2019307,XIAO201915} combined the POD-RBF method with the domain decomposition method, and it has been proved to be a accurate and efficient ROM for general linear and non-linear flow. Wang et al.\cite{WANG20124827} compared POD-interpolation and POD-projection in a heat transfer problem involving multiple variables. It has proved that the projection is more adaptive for broader heat transfer problems but the interpolation is more accurate when parameter range is relatively narrow. 

The POD method has been widely used to reconstruct the flow field in various engineering problems, but it is rarely used for prediction. Because the POD method is to interpolate the principal components of snapshots, and it is not valid to predict the flow solution outside the parameter space of snapshots. It can not replace the role of the Navier-Stokes equations to calculate the randomly varying unsteady flow field, which is not in the sampled parameter space. For prediction, the POD method is only employed in some fluid-structure interaction problems, such as the periodic oscillation of aircraft wings\cite{2013Reduced}, and the spatial-periodic change of flow field, such as the flow around a cylinder with fixed boundary conditions\cite{2017POD}. 

In this work, we considered the periodic change of boundary conditions and predicted the unsteady flow around three cylinders with a sinusoidal inlet flow. \added[id=A]{The main purpose of this paper is to develop a reduced-order model that significantly reduces computational costs while maintaining high accuracy, specifically for real-time prediction of unsteady flows with periodic boundary conditions. We apply the POD method to obtain a reduced order model of the unsteady incompressible flow problem and then introduce the RBF method to replaced the Galerkin projection method in the online stage, which can greatly improve computing efficiency.}

\added[id=A]{The rest of the paper is organized as follows. Section 2 details the mathematical framework for full-order and reduced-order model construction. Section 3 validates the proposed method via a numerical case of unsteady cylinder flow with time-varying inlet velocity, using 120 snapshots to generate POD modes for long-term prediction and conducting error analysis. Section 4 summarizes the work and draws the main conclusions.}

\section{Mathematical model}
\subsection{Governing equations and discretization}

The construction of ROM is associated with snapshots of the flow field in the offline phase\cite{2019Non}. \replaced[id=E]{To}{In order to} obtain high-precision numerical solutions, we need to build a full-order approximation of the Navier–Stokes equations at first. This work is focused on the unsteady incompressible parametrized Navier–Stokes equations. The mass conservation formula and momentum conservation formula are as follows:

\begin{equation}\label{eq:1}
\left\{\begin{array}{l}{\frac{\partial \boldsymbol u}{\partial t}+(\boldsymbol u \cdot \nabla) \boldsymbol u- \nu \nabla^{2} \boldsymbol u=-\nabla p} \\ {\nabla \cdot \boldsymbol u=0}\end{array}\right. .
\end{equation}

Given three-dimensional calculation domain, $\boldsymbol u \equiv\left(u_{x}, u_{y}, u_{z}\right)^\top$ is the vectorial velocity field, t represents the time, $\nu$ denotes the fluid kinematic viscosity which equals the dynamic viscosity divided by the density $\rho$ , and p denotes the normalized pressure, which equals the pressure divided by the fluid density $\rho$.

The governing equations cannot be solved directly \replaced[id=A]{to obtain the analytical solution}{by the mathematical method}. So we apply the finite volume method(FVM) to discretize respective terms. The flow domain $\Omega$ is divided into a set of finite volumes. Then the governing equation is written for each volume $\Omega_{e}$ in integral form as: 

\begin{equation}\label{eq:2}
\left\{\begin{array}{l}{\int_{\Omega_{e}} \frac{\partial \boldsymbol u}{\partial t} \mathrm{d} \Omega+\int_{\Omega_{e}}  (\boldsymbol u \cdot \nabla) \boldsymbol u \mathrm{d} \Omega-\int_{ \Omega_{e}} \nu \nabla^{2} \boldsymbol u \mathrm{d} \Omega+\int_{ \Omega_{e}} \nabla p \mathrm{d} \Omega=0} \\ {\int_{ \Omega_{e}} \nabla \cdot \boldsymbol  u d \Omega=0}\end{array}\right. .
\end{equation}

By Gauss’s theorem, we can convert volume integrals into closed surface integrals. The equations are discretized as:

\begin{equation}\label{eq:3}
\left\{\begin{array}{l}{\int_{\Omega_{e}} \frac{\partial \boldsymbol u}{\partial t} \mathrm{d} \Omega+\int_{\partial \Omega_{e}} (\boldsymbol n \cdot \boldsymbol u_{f})\boldsymbol u_{f} \mathrm{d} \Gamma-\int_{\partial \Omega_{e}}  \nu \boldsymbol n \cdot \nabla \boldsymbol u \mathrm{d} \Gamma+\int_{\partial \Omega_{e}}  \boldsymbol n p \mathrm{d} \Gamma=0} \\ {\int_{\partial \Omega_{e}} n \cdot \boldsymbol u d \Gamma=0}\end{array}\right. ,
\end{equation}
where $\Gamma$ is the boundary of the flow domain $\Omega$, closed surfaces $\partial \Omega_{e}$ are the boundaries of each finite volume $\Omega_{e}$, $\boldsymbol n$ is the unit normal vector to the surface $\partial \Omega_{e}$ with directing out of the volume. 

Assuming that the velocity and pressure in each volume are uniform and constant, the integration is transformed into the summation of products. We take ${\boldsymbol u}_{e}$ the velocity at the center of the volume to approximate the velocity of the volume. We take $p_{f}$, ${\boldsymbol u}_{f}$ the pressure and velocity at the center of the surface to approximate pressure and velocity at the surface. The equations develop into the following form:

\begin{equation}\label{eq:4}
\left\{\begin{array}{l}{  \frac{\partial {\boldsymbol u}_{e}}{\partial t}  V_{e}+\sum_{f} ( {\boldsymbol S}_{f} \cdot  {\boldsymbol u}_{f} )  {\boldsymbol u}_{f}-\nu \sum_{f}  {\boldsymbol S}_{f} \cdot(\nabla  {\boldsymbol u})_{f}+\sum_{f}  {\boldsymbol S}_{f} p_{f}=0} \\ {\sum_{f}  {\boldsymbol S}_{f} \cdot  {\boldsymbol u}_{f}=0}\end{array}\right. ,
\end{equation}
where ${\boldsymbol S}_{f}$ indicates the surface area vector. \added[id=A]{The face values $\boldsymbol u_{f}$ and $p_{f}$ critical for calculating fluxes at control volume boundaries are derived through linear interpolation from cell-centered values.} The values at the center of the volumes are denoted with the subscript e while values at the center of the faces are denoted with the subscript $f$. 

Stabile et al. \cite{STABILE2018273} made a detailed derivation of the finite volume discretization above. More details can be found in \cite{STABILE2018273}. Set appropriate initial velocity ${\boldsymbol u}_{t=0} $ in $\Omega$ and boundary velocity ${\boldsymbol u}_{\Gamma}$ on the boundary $\Gamma$. The high-precision numerical solutions of the flow field can be obtained. \added[id=A]{In the present case, we employ the PIMPLE algorithm to resolve the coupling problem between pressure and velocity at each time step. This semi-implicit methodology proves particularly effective for transient flow problems by facilitating decoupling between pressure and velocity fields—allowing them to be solved sequentially.}

\subsection{Generation of the POD modes}

We store snapshots at each parametric value at each time step in the form of a matrix. \deleted[id=A]{Suppose that the mesh of the computational domain is generated with $k$ nodes. We can describe the flow field by a $k$-dimensional column vector. If $n$ different values of the flow parameters are set, and each solution involves $m$ time steps, we can get $n \times m$ solution vectors.} The matrices of snapshots are shown below:

\begin{equation}\label{eq:5}
	{A}_{\boldsymbol u}=\left[\begin{array}{cccc}
		{\boldsymbol u}_{1}(bc_{1},t_{1}) & {\boldsymbol u}_{1}(bc_{2},t_{2}) & \cdots & {\boldsymbol u}_{1}(bc_{n},t_{m}) \\
		{\boldsymbol u}_{2}(bc_{1},t_{1}) & {\boldsymbol u}_{2}(bc_{2},t_{2}) & \cdots & {\boldsymbol u}_{2}(bc_{n},t_{m}) \\
		\vdots & \vdots &  &  \vdots\\
		{\boldsymbol u}_{k}(bc_{1},t_{1}) & {\boldsymbol u}_{k}(bc_{2},t_{2}) & \cdots & {\boldsymbol u}_{k}(bc_{n},t_{m})
	\end{array}\right]
	{\ \in R^{k\times (n\times m)\added{\times 3}}},
\end{equation}

\begin{equation}\label{eq:6}
	{A}_{p}=\left[\begin{array}{cccc}
		p_{1}(bc_{1},t_{1}) & p_{1}(bc_{2},t_{2}) & \cdots & p_{1}(bc_{n},t_{m}) \\
		p_{2}(bc_{1},t_{1}) & p_{2}(bc_{2},t_{2}) & \cdots & p_{2}(bc_{n},t_{m}) \\
		\vdots & \vdots &  &  \vdots\\
		p_{k}(bc_{1},t_{1}) & p_{k}(bc_{2},t_{2}) & \cdots & p_{k}(bc_{n},t_{m})
	\end{array}\right]
	{\ \in R^{k\times (n\times m)} },
\end{equation}
where $A_{\boldsymbol u}$ is the snapshots matrix of the velocity field, and $A_{p}$ is the snapshots matrix of the pressure field. \replaced[id=A]{Each column in the matrices $A_{\boldsymbol u}$ and $A_{\boldsymbol p}$ corresponds to a snapshot with a specific boundary condition at a specific time step. Here, $n$ denotes the total number of different boundary conditions applied in the simulations while $m$ represents the total number of time steps considered in each simulation. The rows of these matrices, which are $k$ in number, correspond to the spatial nodes, or grid points.}{Vector $\boldsymbol x_{i} (i=1,2,...,k)$ is the value of flow parameters used as the boundary condition. $t_{i} (i=1,2,...,m)$  indicates time steps. Both matrices have $k$ rows and $n \times m$ columns. }

The snapshots matrix of velocity $A_{\boldsymbol u}$ is taken as an example. We need to find a set of ordered orthogonal modes by which all the solutions in the snapshot's space can be described optimally. That is to say, we need to find a set of $\boldsymbol \phi_{i} (i=1,2,...,r)$ to minimize the mean square error between the snapshots and their reduced-order approximation:

\begin{equation}\label{eq:7}
	E= \frac{1}{r} \sum_{i=1}^{r}\left\| \boldsymbol{u}_{i}(bc,t)-\widehat{\boldsymbol{u}}_{i}(bc,t) \right\|_{L^{2}}^{2},
\end{equation}

\begin{equation}\label{eq:8}
	\widehat{\boldsymbol u} (bc,t)= \sum_{i=1}^{r} \alpha_{i}(bc,t) \boldsymbol \phi_{i}, 
\end{equation}
where $n \times m$ is the number of considered snapshots. $\boldsymbol u_{i}(bc,t)$ is the high-fidelity solution of velocity and $\widehat{\boldsymbol{u}}_{i}(bc,t)$ is the reduced-order approximation. r denotes the number of POD modes. $\alpha_{i}(bc,t)$ is the coefficients of POD modes.

We can obtain the orthonormal modes by applying SVD on the snapshots matrix as Eq. \eqref{svd}:

\begin{equation}\label{svd}
A_{\boldsymbol u}=Q_{\boldsymbol u} \Sigma V_{\boldsymbol u}^{\top}=Q_{\boldsymbol u}\left(\begin{array}{ll}{D} & {0} \\ {0} & {0}\end{array}\right) V_{\boldsymbol u}^{\top} .
\end{equation}

Among them, $Q_{\boldsymbol u} \in R^{k\times k}$ and $V_{\boldsymbol u} \in R^{(n\times m) \times (n\times m)}$ are both orthogonal matrices. The matrix $D$ is a diagonal matrix. The elements on the diagonal are singular values, which indicate how much information the corresponding POD mode contains. 

\replaced[id=A]{In snapshot-based flow field, there are far more spatial discrete points than time discrete points, that is, the number of rows $k$ is far greater than the number of columns $n \times m$ in the matrix $A_{\boldsymbol u}$. The computer will consume more costs, whether it is storing, or calculating feature vectors later. Therefore, to solve this problem, the correlation matrix is calculated first to obtain a matrix with $n \times m$ rows and $n \times m$ columns, thus reducing the computation amount.}{However, it was proved to be computationally expensive and make the dimension of the grid increase by Stabile et al. \cite{2017POD}. Therefore, we choose the equivalent and more efficient POD method instead. At first, we compute the correlation matrix of the velocity field snapshots:} 

\begin{equation}\label{eq:10}
	(A_{\boldsymbol u}^{\top}, A_{\boldsymbol  u})_{L_{2}}=\left[\begin{array}{ccc}{\left( \boldsymbol u_{1}, \boldsymbol u_{1}\right)_{L_{2}}} & {\left( \boldsymbol u_{1}, \boldsymbol u_{2}\right)_{L_{2}}} & {\dots\left( \boldsymbol u_{1}, \boldsymbol u_{n \times m}\right)_{L_{2}}} \\ {\left( \boldsymbol u_{2}, \boldsymbol u_{1}\right)_{L_{2}}} & {\left( \boldsymbol u_{2}, \boldsymbol u_{2}\right)_{L_{2}}} & {\cdots\left( \boldsymbol u_{2}, \boldsymbol u_{n \times m}\right)_{L_{2}}} \\ {\vdots} & {\vdots} & {\vdots} \\ {\left( \boldsymbol u_{n \times m}, \boldsymbol u_{1}\right)_{L_{2}}} & {\left( \boldsymbol u_{n \times m}, \boldsymbol u_{2}\right)_{L_{2}}} & {\cdots\left( \boldsymbol u_{n \times m}, \boldsymbol u_{n \times m}\right)_{L_{2}}}\end{array}\right].
\end{equation}

The correlation matrix is decomposed by the eigen orthogonalization method. Then the orthogonal matrix $V_{\boldsymbol u}$ and eigenvalue matrix $\Sigma$ are obtained.

\begin{equation}\label{eq:11}
	(A_{\boldsymbol u}^{\top}, A_{\boldsymbol u})_{L_{2}}=V_{\boldsymbol u}\left[\begin{array}{cccc}{\lambda_{1}^{\boldsymbol u}} & {0} & {\cdots} & {0} \\ {0} & {\lambda_{2}^{\boldsymbol u}} & {\cdots} & {0} \\ {0} & {0} & {\cdots} & {\lambda_{n \times m}^{\boldsymbol u}}\end{array}\right] V_{\boldsymbol u}^{\top},
\end{equation}
where ${V}_{\boldsymbol u}$ is an orthogonal square matrix whose columns are the POD modes. $\lambda_{i}^{\boldsymbol u} (i=1,2,...,n \times m)$ are the eigenvalues in descending order. Then we can get a sets of  $\boldsymbol \phi_{i}$ as follows:

\begin{equation}\label{eq:12}
	\boldsymbol \phi_{i}=\frac{1}{\sqrt{\lambda_{i}^{\boldsymbol u}}} A_{\boldsymbol u} (V_{\boldsymbol u})_{i} =\frac{1}{\sqrt{\lambda_{i}^{\boldsymbol u}}} \sum_{j=1}^{n \times m} {\boldsymbol u}_{j} (V_{\boldsymbol u})_{i j}.
\end{equation}

The POD modes of pressure field $\psi_{i}$  (i = 1,2,...,$n \times m$) can be obtained \replaced[id=E]{similarly}{with the same process, then we can obtain}:

\begin{equation}\label{eq:13}
	\psi_{i}=\frac{1}{\sqrt{\lambda_{i}^{p}}} A_{p} (V_{p})_{i} =\frac{1}{\sqrt{\lambda_{i}^{p}}} \sum_{j=1}^{n \times m} p_{j} (V_{p})_{i j},
\end{equation}
where $(V_{\boldsymbol u})_{i}$ is the \added{$i$}-th row vector of matrix $V_{\boldsymbol u}$, $\boldsymbol u_{j}$ is the \added{$j$}-th snapshot of velocity field which is the \added{$j$}-th column of matrix $A_{\boldsymbol u}$, $(V_{\boldsymbol u})_{i j}$ is the element in row \added{$i$} and column \added{$j$} of the matrix $V_{\boldsymbol u}$, $ (V_{p})_{i}$ is the \added{$i$}-th row vector of matrix $V_{p}$,  $p_{j}$ is the \added{$j$}-th snapshot of pressure field which is the \added{$j$}-th column of matrix $A_{p}$, $(V_{p})_{i j}$ is the element in row \added{$i$} and column \added{$j$} of the matrix $ V_{p}$.

The flow field can be reconstructed quickly by a linear combination of POD modes. \replaced[id=A]{Eigenvalues serve as indicators of the significance or energy associated with their corresponding modes. Modes linked to larger eigenvalues encapsulate the main flow characteristics, whereas those tied to smaller eigenvalues tend to represent lesser energy or noise. By preserving the largest eigenvalues along with their respective modes, we can effectively capture the most critical information present in the dataset. Consequently, based on these eigenvalue assessments, an appropriate reduction in the number of POD modes can be achieved. It is}{According to the eigenvalue, the number of POD modes can be appropriately reduced. The mode with a higher eigenvalue is selected because the eigenvalue represents the information content of each mode.} proved that these POD modes can minimize the average of the error between the snapshots and their reduced-order approximation \cite{LIANG2002527}.

The velocity field and pressure field can be expressed optimally using the selected first r POD modes:

\begin{equation}\label{eq:14}
	\boldsymbol u (bc,t)\approx \boldsymbol u_{ROM} (bc,t)=\sum_{i=1}^{r} \alpha_{i}(bc,t) \boldsymbol \phi_{i},
\end{equation}

\begin{equation}\label{eq:15}
	p (bc,t)\approx p_{ROM} (bc,t)=\sum_{i=1}^{r} \beta_{i}(bc,t) \psi_{i}.
\end{equation}

By finding the temporal coefficients $\alpha_{i}(bc,t)$ and $ \beta_{i}(bc,t)$, we can quickly obtain the reduced-order approximate of the high-precision flow field.  

\subsection{RBF Interpolation method}

\added[id=A]{The Radial Basis Function (RBF) method is a popular technique used in various fields of computational mathematics and engineering, particularly for solving boundary value problems \cite{JANKOWSKA2024205, KARAGEORGHIS2022196} and interpolation tasks \cite{DOBRAVEC202277, LIN2024178}. RBFs are known for their high accuracy in approximating functions and interpolating scattered data in multi-dimensional spaces.} 

Typical RBFs include Gaussian, multi-quadric, inverse quadratic, inverse multi-quadric, polyharmonic spline, thin plate spline, \replaced[id=E]{linear, cubic, and Matérn functions}{etc}. \replaced[id=R1]{Mathematically, RBF can be expressed as: $\Theta_i(\boldsymbol x)=\theta \left(\left\|\boldsymbol x- \boldsymbol x_{i}\right\|\right)$, where $\boldsymbol x$ represents any point in space, $\boldsymbol x_{i}$ denotes the center of the RBF, $\theta$ indicates the specific form of the RBF, and $\left\|\boldsymbol x- \boldsymbol x_{i}\right\|$ indicates the Euclidean distance from point $\boldsymbol x$ to center $\boldsymbol x_{i}$. In this work, we adopt the Gaussian radial basis function $\exp(-(\varepsilon  \left\|\boldsymbol x- \boldsymbol x_{i}\right\|)^2)$ for all interpolation tasks due to its excellent smoothness properties. A shape parameter of $\varepsilon=1.0$ was used, which is a standard value commonly employed in similar POD-RBF studies and demonstrated consistent stable performance in our preliminary numerical tests.}{RBF, expressed as $\Theta\left(\left\|x- x_{i}\right\|\right)$, is a scalar function symmetric along the radial direction, whose function value depends only on the radial distance (usually Euclidean distance) between the sample point to the center point. Therefore, it is very suitable for high-dimensional interpolation. With increasing the dimension of interpolation parameters, the complexity will not be increased.}

RBF is introduced to determine the POD coefficients $\alpha_{i}(bc,t)$ and $ \beta_{i}(bc,t)$ in Eq. \eqref{eq:14} and Eq. \eqref{eq:15}. It is essentially a nonlinear regression method. After determining POD modes, each original snapshot can be represented by a set of coefficients. The snapshots matrix can be represented by the coefficient matrix, whose amount of data is much less than the snapshots matrix. The coefficient matrix is composed of the POD coefficients of all snapshots to all POD modes.

There are two application cases to obtain ROM based on RBF functions. One is taking boundary condition parameters as the interpolation variable, which is suitable for steady flow. The other is to take the velocity and pressure field as the interpolation variable at each step, which is suitable for unsteady flow. Next, the two interpolation ways are introduced, respectively.

In the context of steady flow, the flow field is determined by the initial boundary conditions and remains invariant over time. Consequently, we can establish a response relationship between these boundary conditions and the time-independent POD coefficients $\alpha_{i}(bc)$ using RBFs as follows:

\begin{equation}\label{eq:16}
	\left[\begin{array}{c}
		\alpha_1\\
		\alpha_2 \\
		\vdots \\
		\alpha_r
	\end{array}\right](bc)=\left[\begin{array}{cccc}
		\omega_{11} &\omega_{12} & \cdots & \omega_{1n}\\
		\omega_{21} & \omega_{22} & \cdots & \omega_{2n} \\
		\vdots & \vdots &  &  \vdots\\
		\omega_{r1} & \omega_{r2} & \cdots & \omega_{rn}
	\end{array}\right]
	\left[\begin{array}{c}
		\Theta_1\\
		\Theta_2 \\
		\vdots \\
		\Theta_n
	\end{array}\right](bc) ,
\end{equation}
where $\alpha_{i} (i=1,2,...,r)$ denotes the coefficient of the $i$-th POD modes, $bc$ denotes the input boudary condition, $\omega_{ij}$ denotes the weight of the RBF $\Theta_j$ in the interpolation formula of $\alpha_{i}$, $\Theta_{j}(bc) = \theta(\|bc-bc_{j}\|)(j = 1,2,...,n)$ indicates RBF with center point $bc_j$. Through \deleted[id=E]{the} POD \deleted[id=E]{process}, we get the coefficients of the $r$ POD modes under $n$ different boundary conditions. Each boundary condition corresponds to a unique set of coefficients. Substituting them into the above formulas, we can obtain all $\omega_{ij}$. For each POD mode, we can fit the interpolation formula as follows:

\begin{equation}\label{eq:17}
	\alpha_{i}(bc)=\sum_{j=1}^{n} \omega_{ij} \Theta_j (bc)  , \ (i=1,2,...,r).
\end{equation}

In the same way, we can construct formulas to interpolate the coefficients of pressure modes $\beta_{i}(\boldsymbol {x})$. For any new boundary condition $bc', bc' \notin {bc_{1},bc_{2},...bc_{n}}$, the coefficients of the corresponding POD modes can be calculated by the above formulas, which is easier and more efficient than the Galerkin-projection method.

The above \replaced[id=E]{applies to}{process is for} steady-state flow, where the RBF model is used to fit the one-to-one correspondence between the boundary conditions and the flow field. For unsteady flow, the Navier-Stokes equations involve the first-order time derivatives of physical quantities. Assuming that, for an incompressible unsteady flow,  the flow can be simplified as a Markov process when the time interval is short enough. \added[id=A]{Meyer et al. \cite{PhysRevE} discusses the testing of the Markov hypothesis for fluid and inertial particles in turbulence and fluid particles, proving that fluid flow can be regarded as a Markov process for certain scales and conditions. In CFD of unsteady flow, the numerical methods employed is time-stepping algorithms. This progression can be seen as analogous to a Markov process, where the next state of the flow field is determined solely by its current state, reflecting the memoryless property of Markov chains.} The velocity and pressure at the current time step only depend on that at the previous time step. \replaced[id=A]{We can establish the RBF model in time, by replacing the input boundary condition $bc$ of Eq. \eqref{eq:16} with reduced coefficents $\boldsymbol  c$, and the output is the reduced coefficients at the next step:}{List the equations with reference to Eq. \eqref{eq:16} and replace the input variable $bc$ with $\boldsymbol  c$ in each RBF. $\boldsymbol c=\left[\begin{array}{l} \boldsymbol \alpha  \  \boldsymbol \beta  \end{array}\right]^{\top}$ is a complete set of POD coefficients for both velocity and pressure fields at each time step.}

\begin{equation}\label{eq:18}
	\left[\begin{array}{c}
		f_1\\
		f_2 \\
		\vdots \\
		f_r
	\end{array}\right](\boldsymbol c)=\left[\begin{array}{cccc}
		\omega_{11} &\omega_{12} & \cdots & \omega_{1m}\\
		\omega_{21} & \omega_{22} & \cdots & \omega_{2m} \\
		\vdots & \vdots &  &  \vdots\\
		\omega_{r1} & \omega_{r2} & \cdots & \omega_{rm}
	\end{array}\right]
	\left[\begin{array}{c}
		\Theta_1\\
		\Theta_2 \\
		\vdots \\
		\Theta_m
	\end{array}\right](\boldsymbol c) ,
\end{equation}
where $\boldsymbol c=\left[\begin{array}{l} \boldsymbol \alpha  \  \boldsymbol \beta  \end{array}\right]^{\top}$ is a complete set of POD coefficients for both velocity and pressure fields at current timestep. $f_{i}(\boldsymbol c)$ (i=1,2,...,r) is the coeffient of $i$-th POD modes at next timestep. r is the number of all POD modes (velocity and pressure). $\boldsymbol c^{t_{i}}$ (i=1,2,...,m), the center of each RBF, is complete POD coeffients of each snapshot at each timestep.  All the fomulas can be written in the general form.

\begin{equation}\label{eq:19}
	f_{i}(\boldsymbol c)=\sum_{j=1}^{m} \omega_{ij} \Theta\left(\left\|\boldsymbol c-\boldsymbol c^{t_{j}}\right\|\right)  , \ (i=1,2,...,r).
\end{equation}

\begin{figure}[!htbp]
	\centering
	\includegraphics[width=12cm]{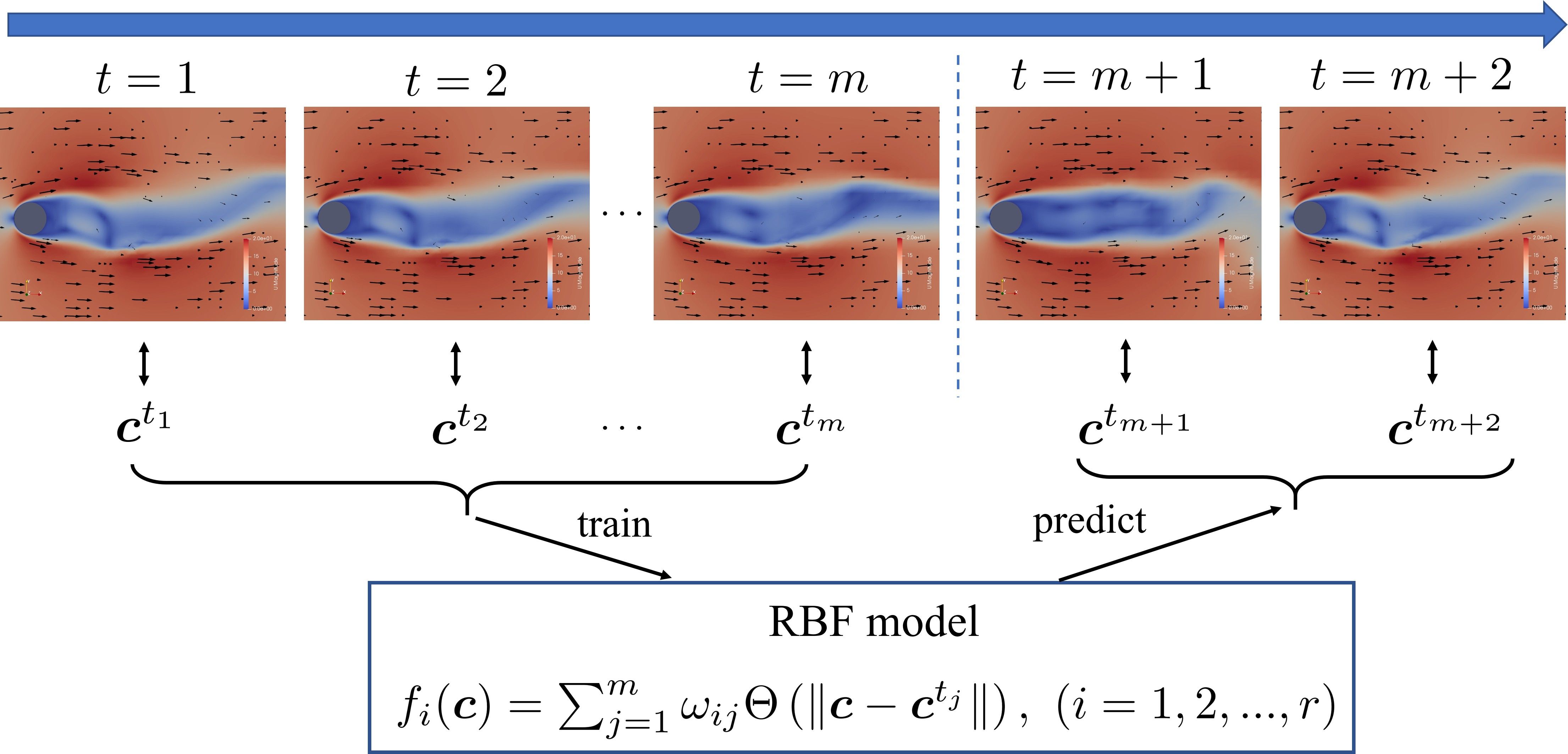}
	\caption{The RBF model of unsteady flow architecture}
	\label{fig:RBFinTime}
\end{figure}

Substituting the coefficient $\boldsymbol c^{t_{j}}= \left[\begin{array}{l} (\boldsymbol u^{t_{j}},\boldsymbol \phi) \\  (p^{t_{j}},\boldsymbol \psi)\end{array}\right]$ at each time $t_{j}$  and the function value $f_{i}(\boldsymbol c^{t_{j}})= c_{i}^{t_{j+1}}$ (j=1,2,...,m) into Eq. \eqref{eq:19} to calculate the weight $\omega_{ij}$ of each RBF. Then we can estimate the POD coefficients of the current time step by that of the last time step using the RBF interpolation and obtain the solution of velocity and pressure by a linear combination of POD modes. \added[id=A]{Upon the completion of the training, the RBF model can promptly predict the reduced coefficients of flow filed at the next moment when the reduced coefficients at any given moment is input. Therefore, we can obtain the reduced-order coefficient of the flow field at each subsequent moment step by step subsequently, as long as we obtain the reduced-order coefficients of the initial flow field: $c_{t=1} \stackrel{f}{\longrightarrow}c_{t=2},c_{t=2} \stackrel{f}{\longrightarrow}c_{t=3},\cdots,c_{t=m} \stackrel{f}{\longrightarrow}c_{t=m+1},\cdots$, as is shown in Fig. \ref{fig:RBFinTime}.}

\begin{algorithm}
	\caption{Reduced-order modelling based on POD and RBF}
	\label{alg1}
	\small
	\begin{algorithmic}
		\REQUIRE Time dependent boundary conditions $\boldsymbol x(t)$and time step $n$;
		\ENSURE Velocity field interpolation hypersurface:$f^{u}_{1},f^{u}_{2},...f^{u}_{r}$ and ressure field interpolation hypersurface:$f^{p}_{1},f^{p}_{2},...f^{p}_{r}$;
		\STATE 1.The snapshot data of velocity/pressure obtained by the full-order model
		: $[\boldsymbol S^{u}(t=1),\boldsymbol S^{u}(t=2),...,\boldsymbol S^{u}(t=n)]$ and $[\boldsymbol S^{p}(t=1),\boldsymbol S^{p}(t=2),...,\boldsymbol S^{p}(t=n)]$;
		\STATE 2.Velocity/pressure modes obtained by POD: ${\boldsymbol \phi_{1},\boldsymbol \phi_{2},...,\boldsymbol \phi_{n}}$ and ${\boldsymbol \psi_{1},\boldsymbol \psi_{2},...,\boldsymbol \psi_{n}}$;
		
		\FOR{$i=1 \to r$}
		\FOR{$j=1 \to n$}
		\STATE Calculate the weight coefficient of each snapshot projected in the POD modes:
		$\alpha_{i j}=\left\langle S^{u}(t=j), \phi_{i}\right\rangle,\beta_{i j}=\left\langle S^{p}(t=j), \psi_{i}\right\rangle$ , $c_{i j}=[\alpha_{i j},\beta_{i j}]$;
		\ENDFOR        
		\STATE $\boldsymbol B^{u}_{i}=[\alpha_{i 1},\alpha_{i 2},...,\alpha_{i n}]^{\top},\boldsymbol B^{p}_{i}=[\beta_{i 1},\beta_{i 2},...,\beta_{i n}]^{\top}$;
		\ENDFOR
		\STATE 5.Obtain $\boldsymbol \omega^{u}$ and $\boldsymbol \omega^{p}$ by solving linear equations:  $\boldsymbol A\boldsymbol \omega = \boldsymbol B$;
		\STATE $\boldsymbol A_{i}\boldsymbol \omega^{u}_{i}=\boldsymbol B^{u}_{i},\boldsymbol A_{i}\boldsymbol \omega^{p}_{i}=\boldsymbol B^{p}_{i}$ ;
		\FOR{$k=1 \to r$}  
		\STATE $f^{u}_{k}(\boldsymbol c)=\boldsymbol \omega^{u}_{k}[\Theta(\left \| \boldsymbol c-\boldsymbol c(t=1)\right \|),\Theta(\left \| \boldsymbol c-\boldsymbol c(t=2)\right \|),...,\Theta(\left \| \boldsymbol c-\boldsymbol c(t=n)\right \|)]$;
		\STATE $f^{p}_{k}(\boldsymbol c)=\boldsymbol \omega^{p}_{k}[\Theta(\left \| \boldsymbol c-\boldsymbol c(t=1)\right \|),\Theta(\left \| \boldsymbol c-\boldsymbol c(t=2)\right \|),...,\Theta(\left \| \boldsymbol c-\boldsymbol c(t=n)\right \|)]$ ;   
		\ENDFOR
		\STATE 6. Reduced-order velocity: $\sum_{i=1}^{r} f^{u}_{i}(\boldsymbol c_{current}) \phi_{i}$, Reduced-order pressure: $\sum_{i=1}^{r} f^{p}_{i}(\boldsymbol c_{current}) \psi_{i}$;	  
	\end{algorithmic}
\end{algorithm}

\section{Numerical Case}

In this section, the proposed method is tested on a numerical case of a three-dimensional unstructured mesh. The unsteady flow around cylinders with time-varying inlet velocity is utilized to demonstrate the efficiency and accuracy of the method.

\subsection{Establishing the ensemble of snapshots}

\begin{figure}[!htbp]
	
	\begin{minipage}{1\textwidth}
		\centering
		\subcaptionbox{Top view of the mesh with dimension and boundaries}
		{\includegraphics[width=9.2cm]{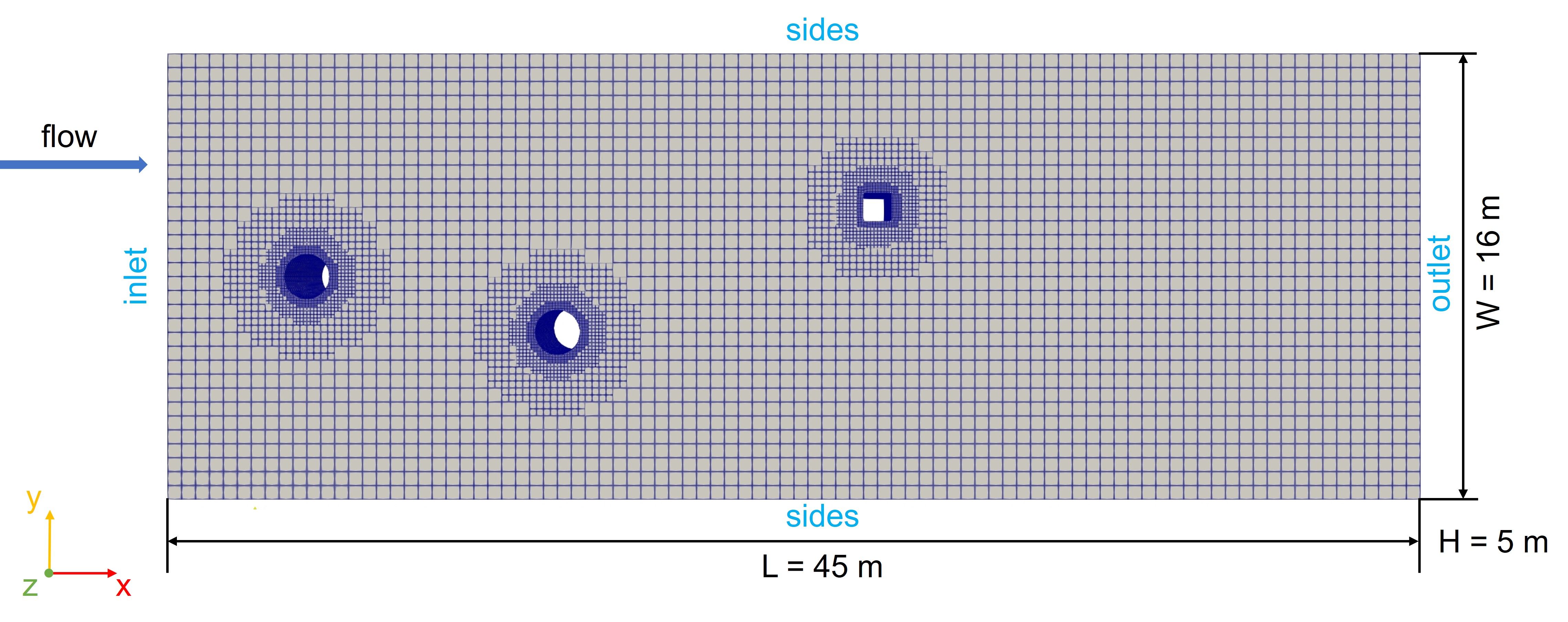}}		
	\end{minipage}	
	
	\begin{minipage}{1\textwidth}
		\centering
		\subcaptionbox{Sectional drawing of cylinders}
		{\includegraphics[width=7cm]{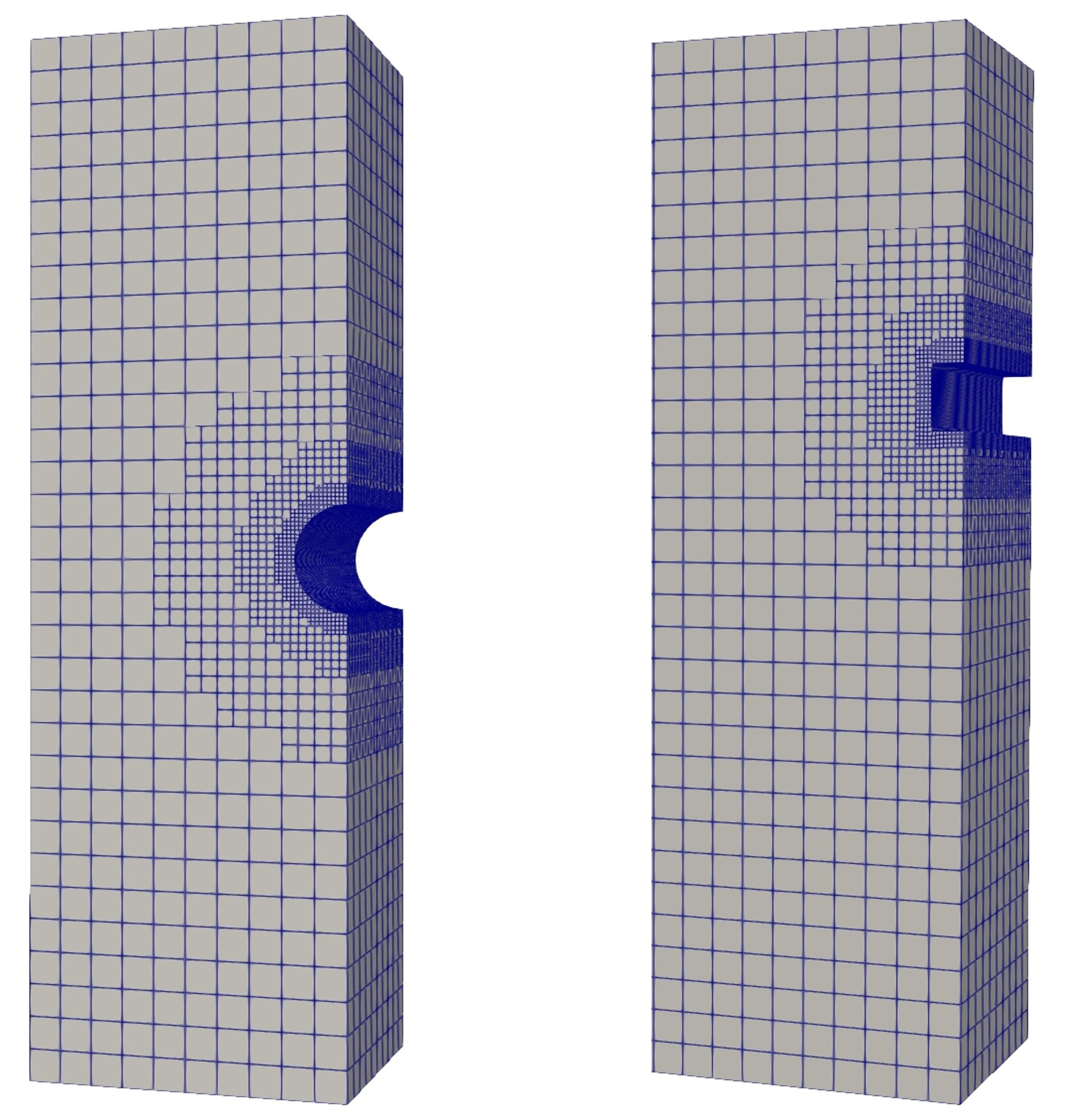}}
	\end{minipage}
	\caption{The computational domain and the unstructured grid mesh}
	\label{fig:grids}
\end{figure}

The simulation is carried on a three-dimensional cuboid domain of length L = 45 m, width W = 16 m, and height H = 5 m. There are two cylinders of radius R = 0.8 m and one square cylinder of length D = 1 m in this domain. For the relatively larger velocity gradient near the cylinders, the mesh was refined in the proximity of each cylinder, as shown in Fig. \ref{fig:grids}. A total of 199650 cells are used.

The left side is the inlet boundary of the computational domain. The velocity of inlet flow is two-dimensional, which is a fixed value $u_{x}$ = 3 m/s in the x-direction and changes as a sinusoidal function $u_{y} = sin(0.5t)$ in the y-direction. The gradient of pressure along the normal direction of the inlet boundary is 0. A slip boundary condition is applied on the front and back sides, where the normal component of velocity is 0 while the tangential component is unchanged. The open boundary condition is applied on the right and the top sides, where the gradient of velocity along the normal direction is 0 and the pressure is 0.

\begin{figure}[!htbp]
	\begin{minipage}{0.5\textwidth}
		\centering
		\subcaptionbox{t = 65s}
		{\includegraphics[width=6cm]{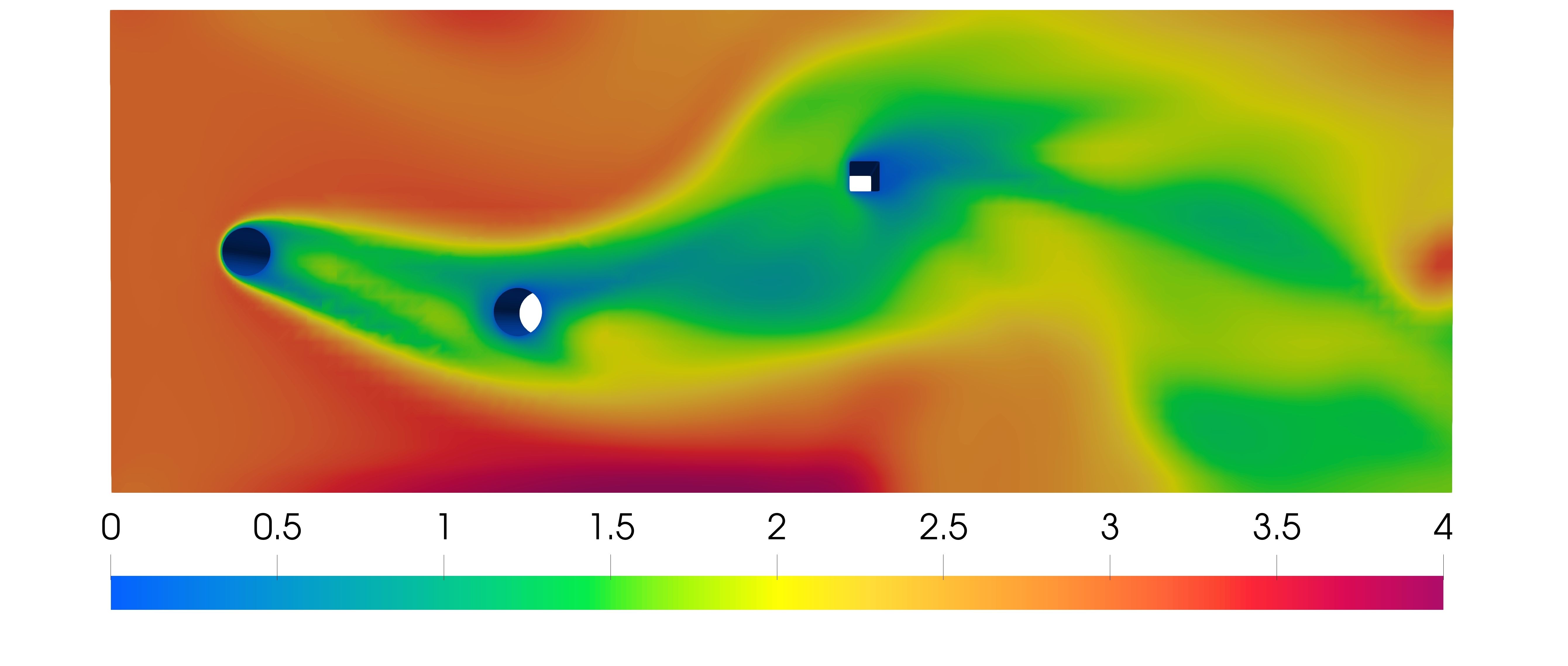}}		
	\end{minipage}
	\begin{minipage}{0.5\textwidth}
		\centering
		\subcaptionbox{t = 90s}
		{\includegraphics[width=6cm]{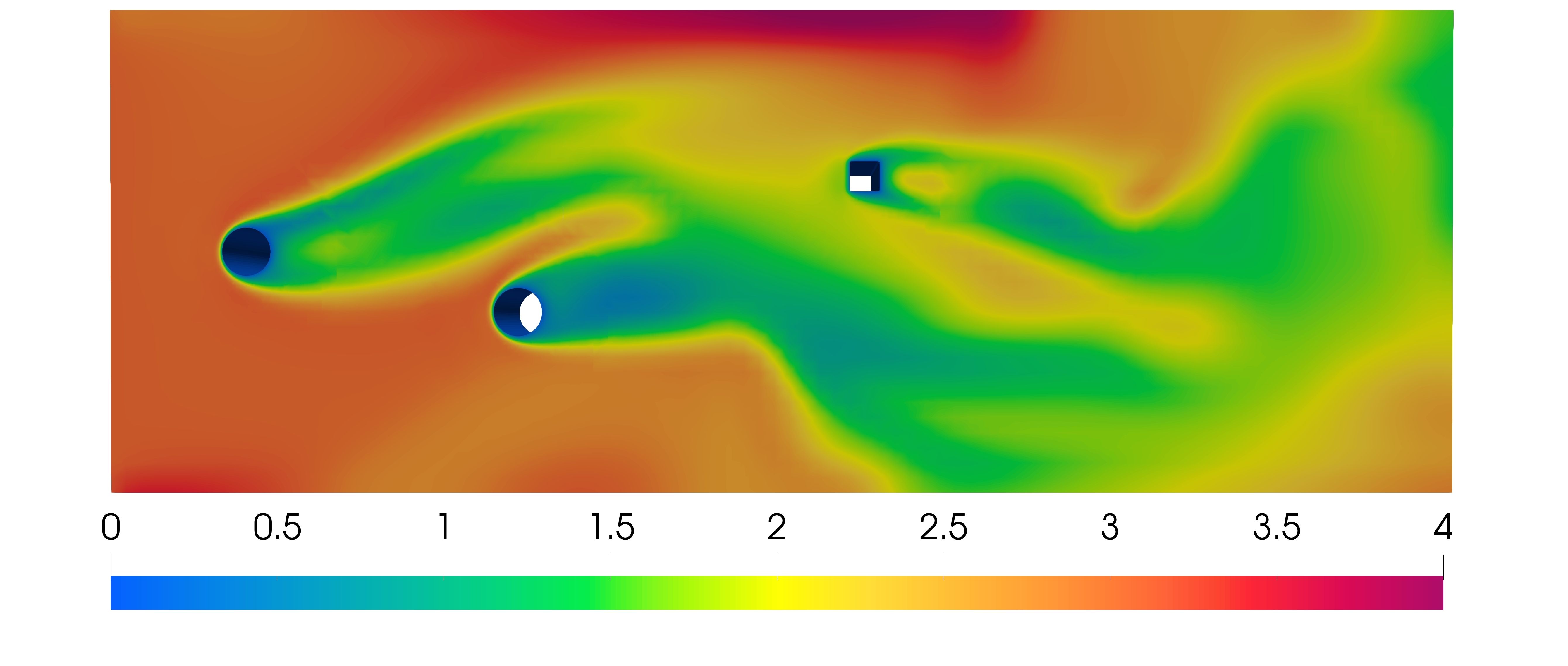}}
	\end{minipage}	
	
	\begin{minipage}{0.5\textwidth}
		\centering
		\subcaptionbox{t = 65s}
		{\includegraphics[width=6cm]{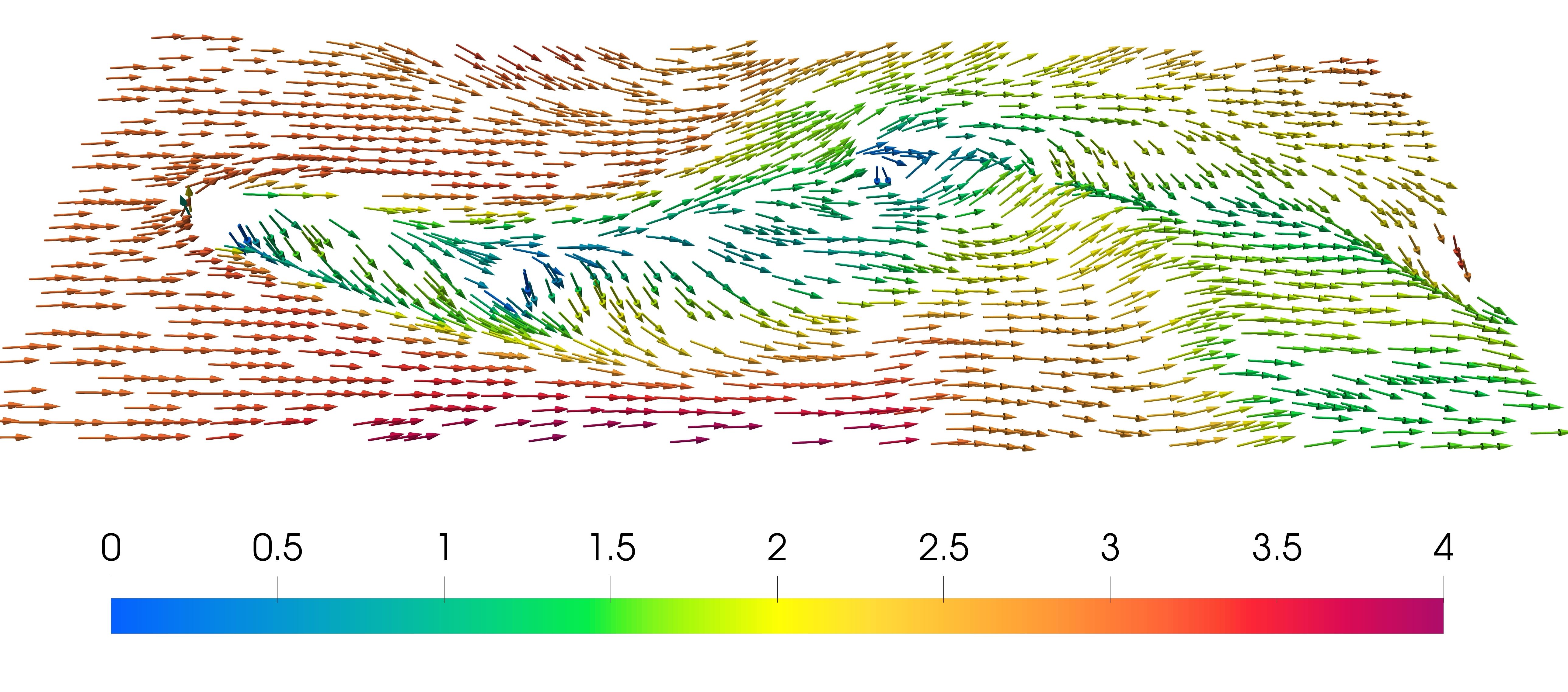}}
	\end{minipage}
	\begin{minipage}{0.5\textwidth}
		\centering
		\subcaptionbox{t = 90s}
		{\includegraphics[width=6cm]{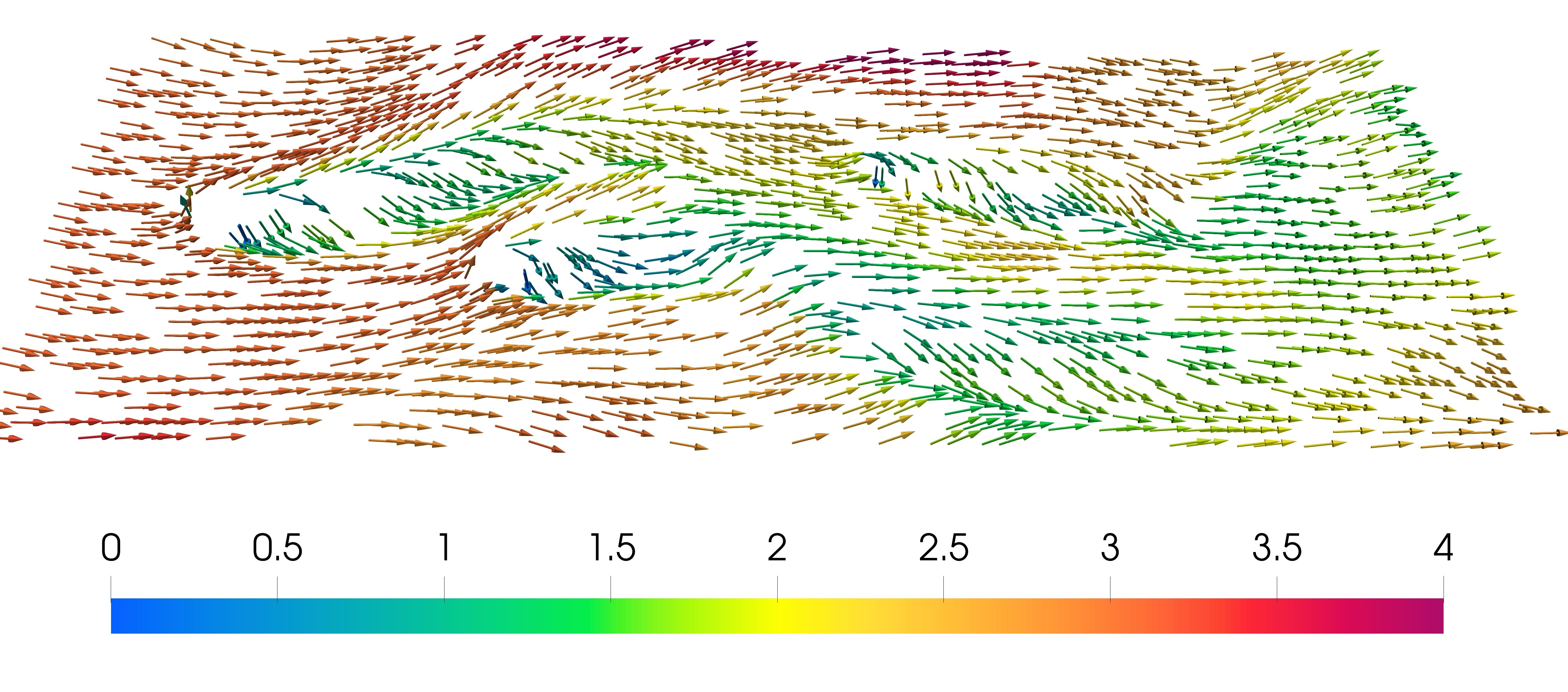}}
	\end{minipage}
	\caption{Snapshots of velocity fields at time steps t=65s and t=90s}
	\label{fig:3snapshots}
\end{figure}

Since we are interested into the correct reconstruction and prediction of the ROM during the periodic inlet velocity, the simulations of the full order model(FOM) are performed during the time period of (30s, 150s] with a time step size of $\Delta t = 0.01s$. 120 snapshots are stored at a regularly spaced time interval of 1s. The snapshot velocity fields and the snapshot pressure fields were used to create snapshot matrices. A column of the matrix contains 199650 node data of each snapshot, and the matrix has 120 columns. For the velocity snapshots matrix, the velocity at each node is three-dimensional and the data amount is about 72 million. Fig. \ref{fig:3snapshots} shows velocity snapshots and the arrow diagram of the horizontal slices at time step t = 65s and t = 90s.

\subsection{Selection of POD modes }

The snapshots obtained during the time period of (30s, 150s] are used to create POD modes and the ROM to predict the solutions during the time period of (150s ,230s]. After applying the POD to the snapshots matrix, we obtain 120 POD modes of velocity and pressure with corresponding eigenvalues. These POD modes provide an optimal representation of the snapshot matrix.

The eigenvalues of POD modes are arranged from high to low, indicating the proportion of information. Table 1 shows the respective eigenvalues and the cumulative eigenvalues of some POD modes.

\begin{table}[h]
	\centering
	\caption{Eigenvalues of the first 10 POD modes and the 20th, 30th modes}
	\begin{tabular}{ccccc}
		\hline
		Mode & $\lambda_{\boldsymbol u}$ & $\lambda_{\boldsymbol u}$(cumulative) & $\lambda_{p}$ & $\lambda_{p}$(cumulative) \\
		\hline
		1     & 0.881176 & 0.881176 & 0.741995 & 0.741995 \\
		2     & 0.049453 & 0.930630 & 0.126293 & 0.868288 \\
		3     & 0.040154 & 0.970783 & 0.048871 & 0.917159 \\
		4     & 0.008454 & 0.979237 & 0.030295 & 0.947454 \\
		5     & 0.007881 & 0.987118 & 0.027080 & 0.974533 \\
		6     & 0.003332 & 0.990450 & 0.005744 & 0.980277 \\
		7     & 0.003263 & 0.993713 & 0.004908 & 0.985185 \\
		8     & 0.001330 & 0.995043 & 0.003119 & 0.988304 \\
		9     & 0.001321 & 0.996364 & 0.003033 & 0.991338 \\
		10    & 0.000551 & 0.996915 & 0.001894 & 0.993232 \\
		20    & 0.000094 & 0.999294 & 0.000168 & 0.998725 \\
		30    & 0.000028 & 0.999789 & 0.000043 & 0.999624\\				
		\hline
	\end{tabular}%
	\label{tab:addlabel}%
\end{table}%

According to the results, we can find the information accounts for less than 1\% from the fourth velocity and the sixth pressure modes with a rapidly decreasing rate. Removing the modes with a small proportion of energy reduces the data dimension, and a part of the information is inevitably lost.

\begin{figure}[!htbp]
	\begin{minipage}{0.32\textwidth}
		\centering
		\subcaptionbox{Mode 1}
		{\includegraphics[width=3.9cm]{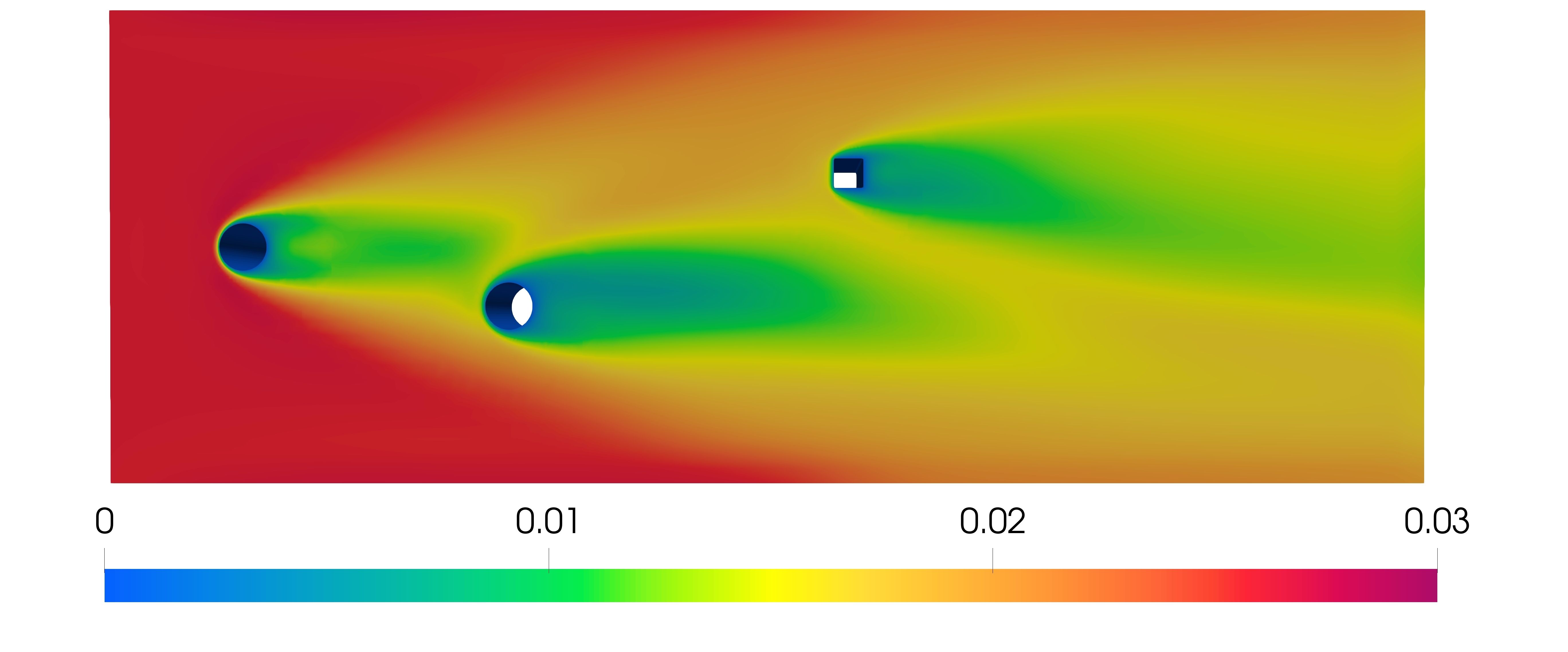}}		
	\end{minipage}
	\begin{minipage}{0.32\textwidth}
		\centering
		\subcaptionbox{Mode 2}
		{\includegraphics[width=3.9cm]{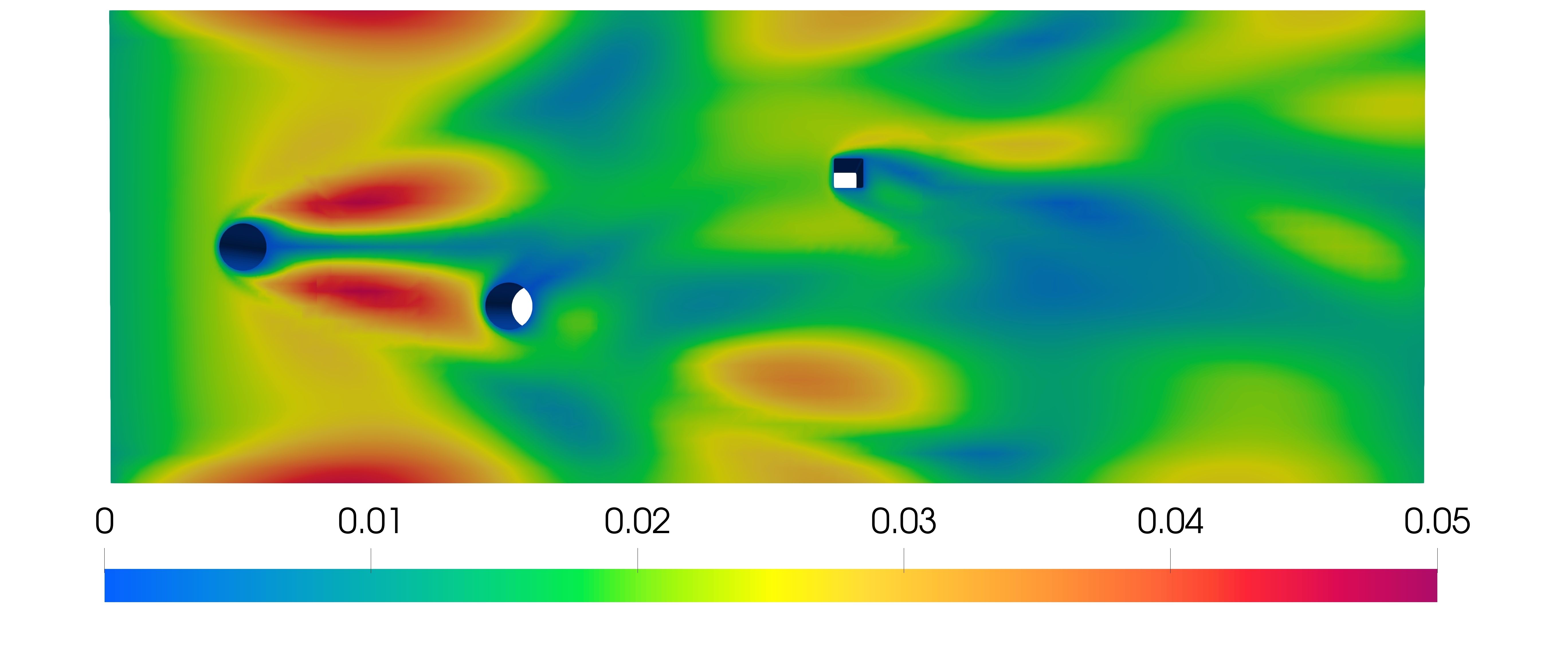}}
	\end{minipage}	
	\begin{minipage}{0.32\textwidth}
		\centering
		\subcaptionbox{Mode 3}
		{\includegraphics[width=3.9cm]{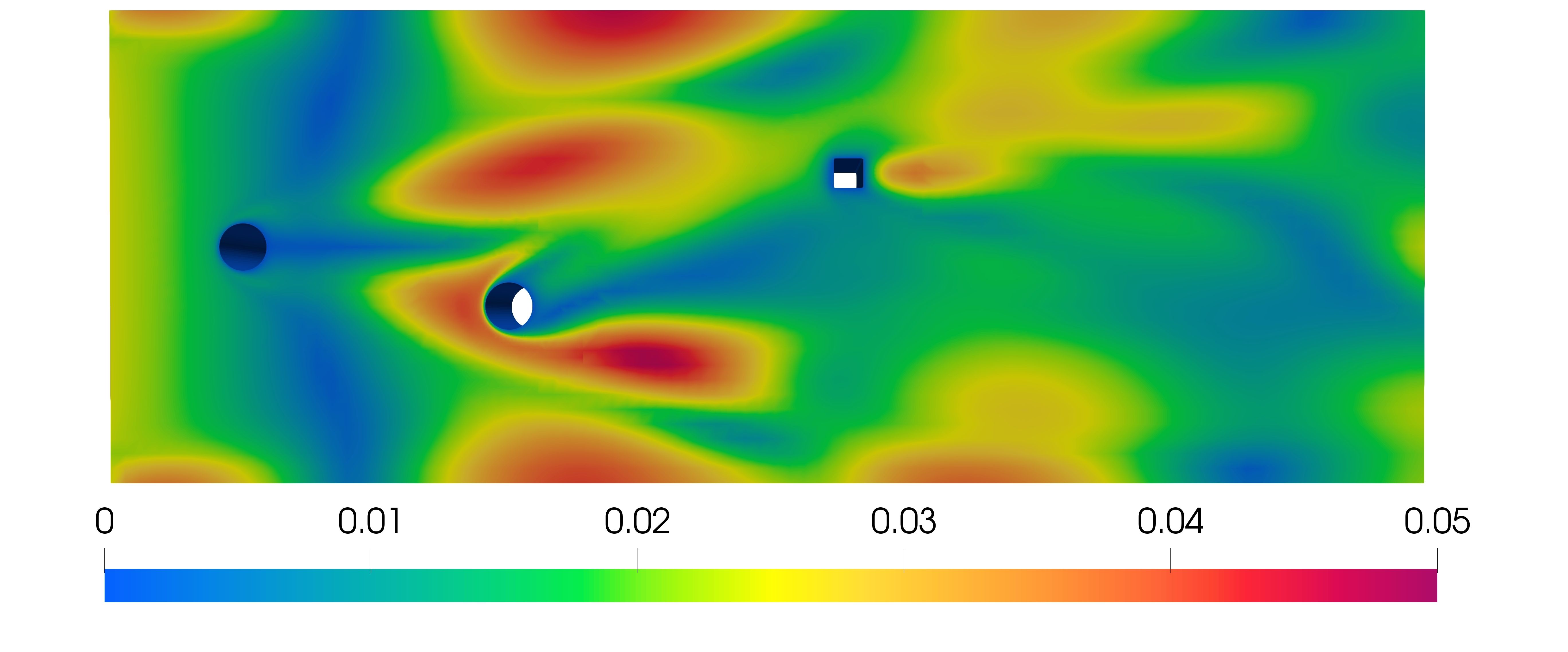}}
	\end{minipage}

	\begin{minipage}{0.32\textwidth}
		\centering
		\subcaptionbox{Mode 4}
		{\includegraphics[width=3.9cm]{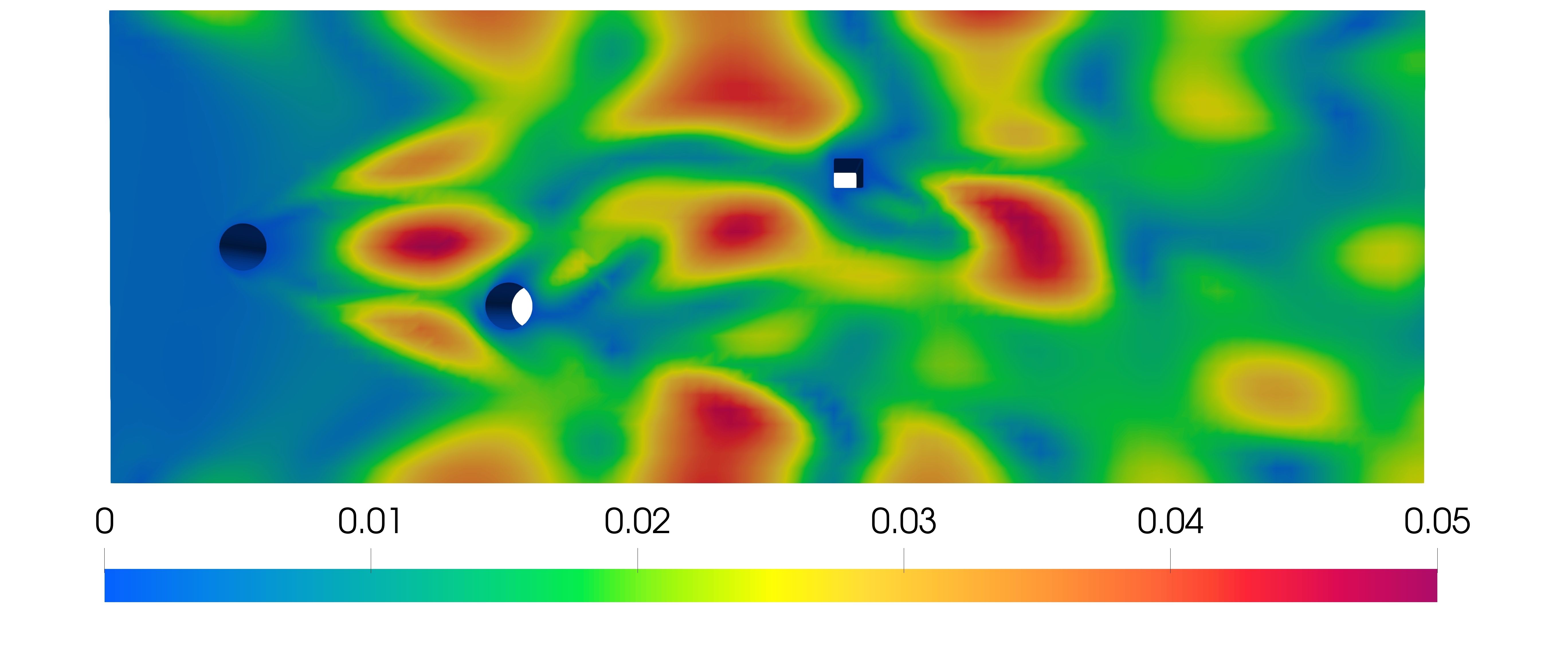}}
	\end{minipage}	
	\begin{minipage}{0.32\textwidth}
		\centering
		\subcaptionbox{Mode 5}
		{\includegraphics[width=3.9cm]{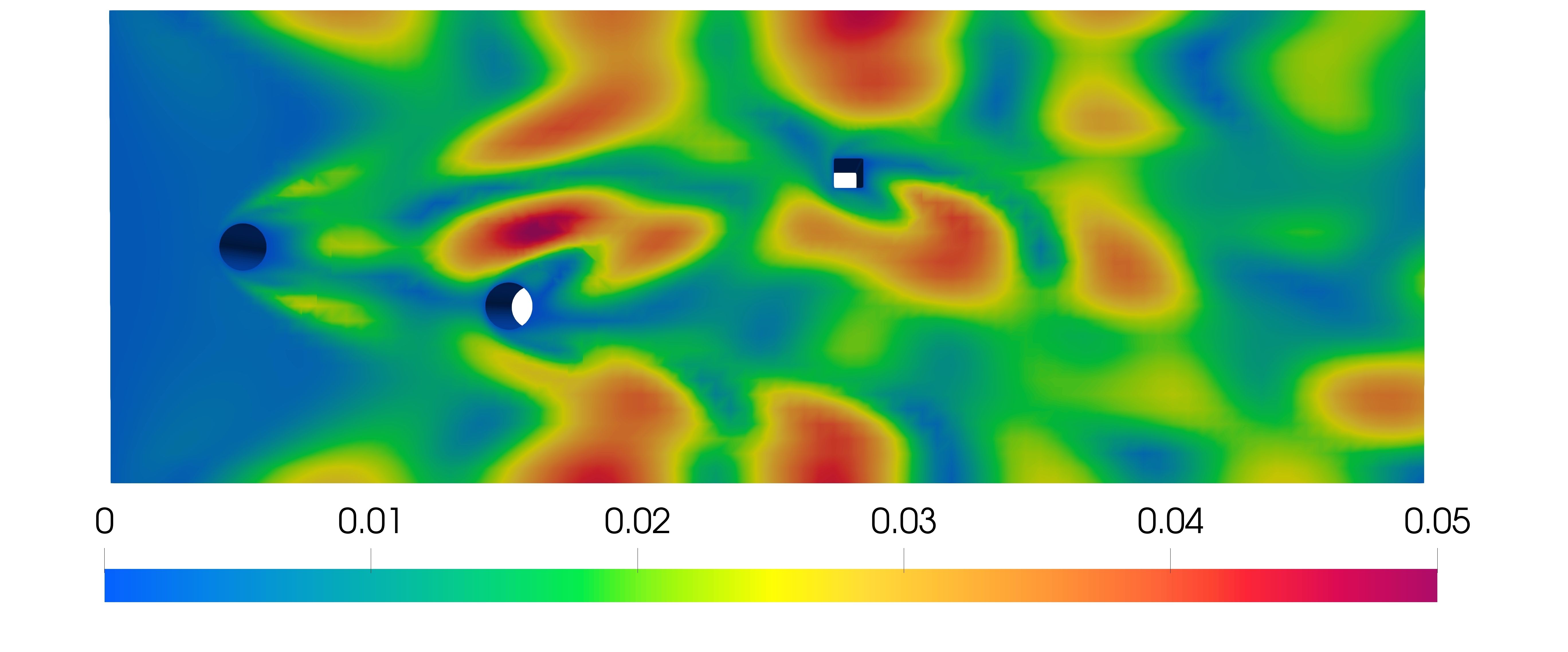}}
	\end{minipage}
	\begin{minipage}{0.32\textwidth}
		\centering
		\subcaptionbox{Mode 6}
		{\includegraphics[width=3.9cm]{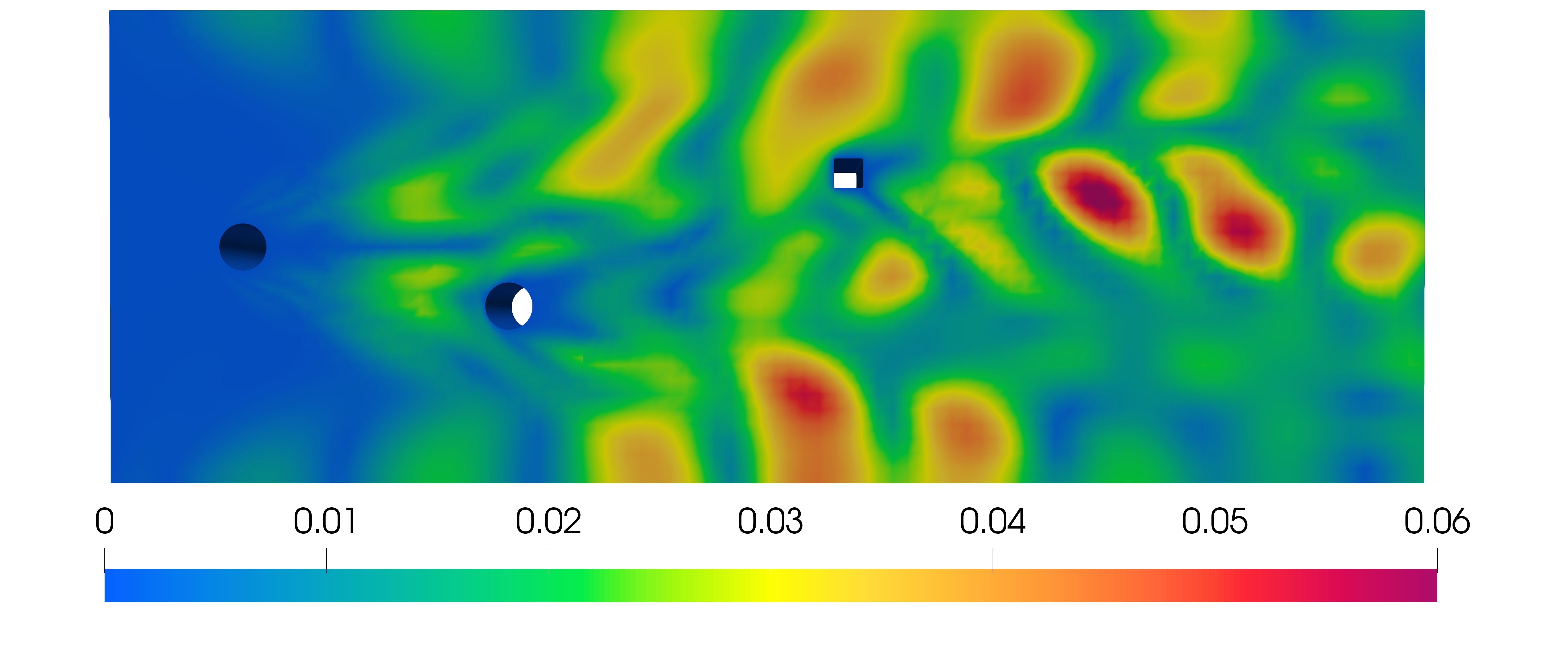}}
	\end{minipage}
	\caption{The first 6 POD modes of velocity fields}
	\label{fig:4modes}
\end{figure}

Assuming that the first 20 modes are retained because of their high energy ratio, the percentage of energy captured can be quantified by the ratio of eigenvalues. Eigenvalues of chosen modes $\sum_{i=1}^{20} \lambda_{i} $ are divided by eigenvalues of all modes $ \sum_{i=1}^{120} \lambda_{i}$ ,which equals 99.93\% for velocity and 99.87\% for pressure. Therefore we use the first 20 POD modes to build the ROM, which causes less than 1\% energy loss. Fig. \ref{fig:4modes} shows the first six POD modes for velocity.

The first mode accounts for the highest energy ratio, more than 88\%. Because each snapshot has a constant inlet flow along the x-direction with fixed value $\boldsymbol u_{x}= 3 m/s$. Modes in front represent the macroscopic, low-frequency and basic flow structure in the computational domain.

\subsection{Reconstruction of velocity fields}

\begin{figure}[!htbp]
	\begin{minipage}{0.5\textwidth}
		\centering
		\subcaptionbox{t = 65s \label{fig:5.1}}
		{\includegraphics[width=6cm]{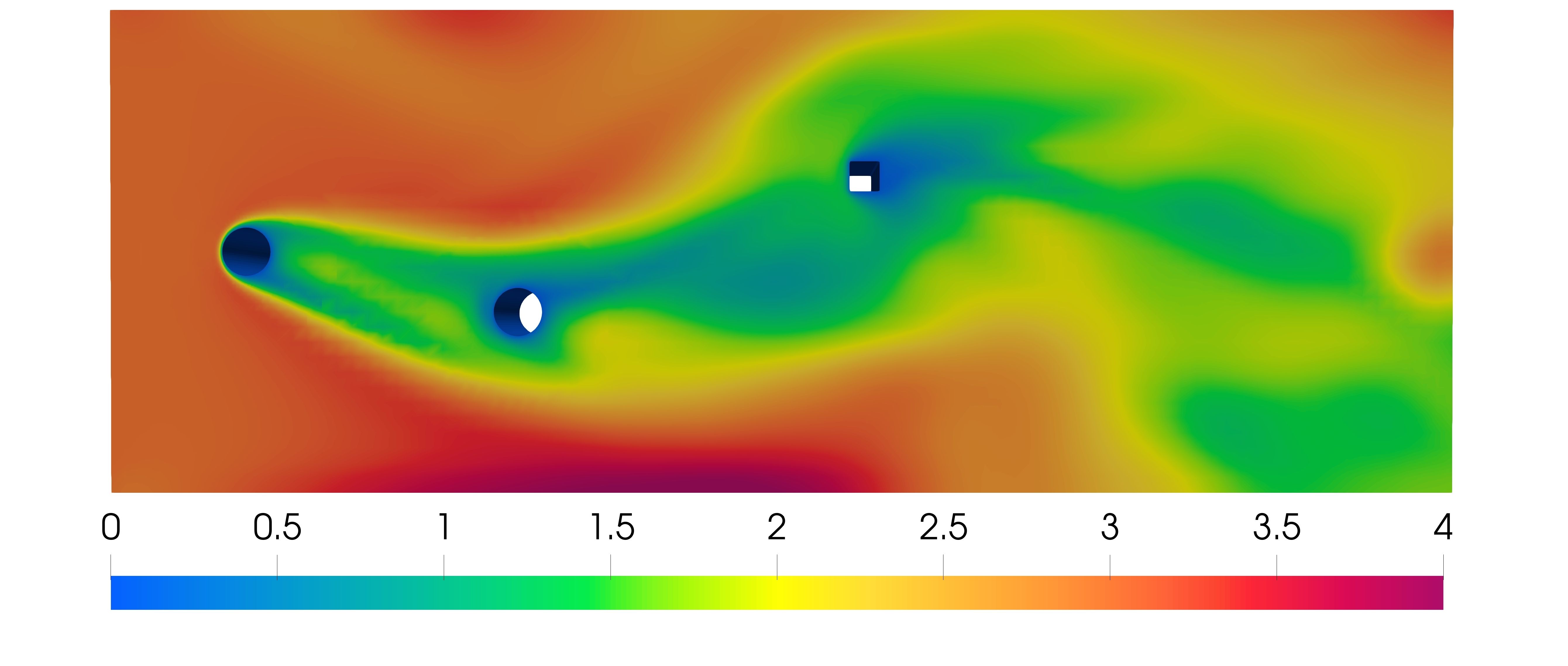}}		
		
	\end{minipage}
	\begin{minipage}{0.5\textwidth}
		\centering
		\subcaptionbox{t = 90s \label{fig:5.2}}
		{\includegraphics[width=6cm]{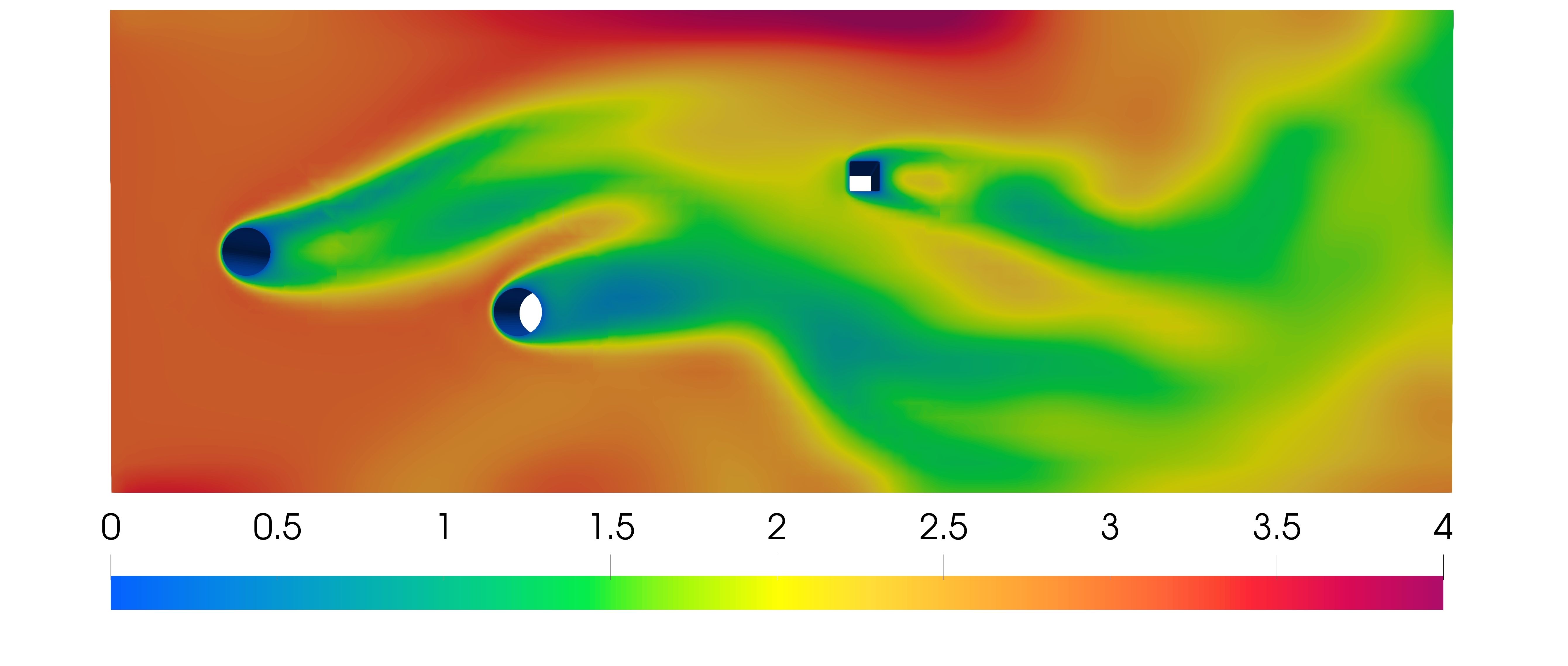}}
	\end{minipage}	
	\caption{velocity fields reconstructed by ROM with 20 POD modes}
	\label{fig:5}
\end{figure}

The snapshots obtained during the time period of (30s, 150s] are projected onto POD modes to obtain 120 sets of POD coefficients, and then the interpolation formula is fitted on that. The coefficient of the initial field $c^{t_{1}}$ is substituted into the interpolation formula as input. All the snapshots are reconstructed by using the formula repeatedly. Fig. \ref{fig:5} shows two reconstruction solutions of the velocity field at time step t=65s and t=90s using the ROM.

\begin{figure}[!htbp]
	\begin{minipage}{0.5\textwidth}
		\centering
		\subcaptionbox{t = 65s \label{fig:65error}}
		{\includegraphics[width=6cm]{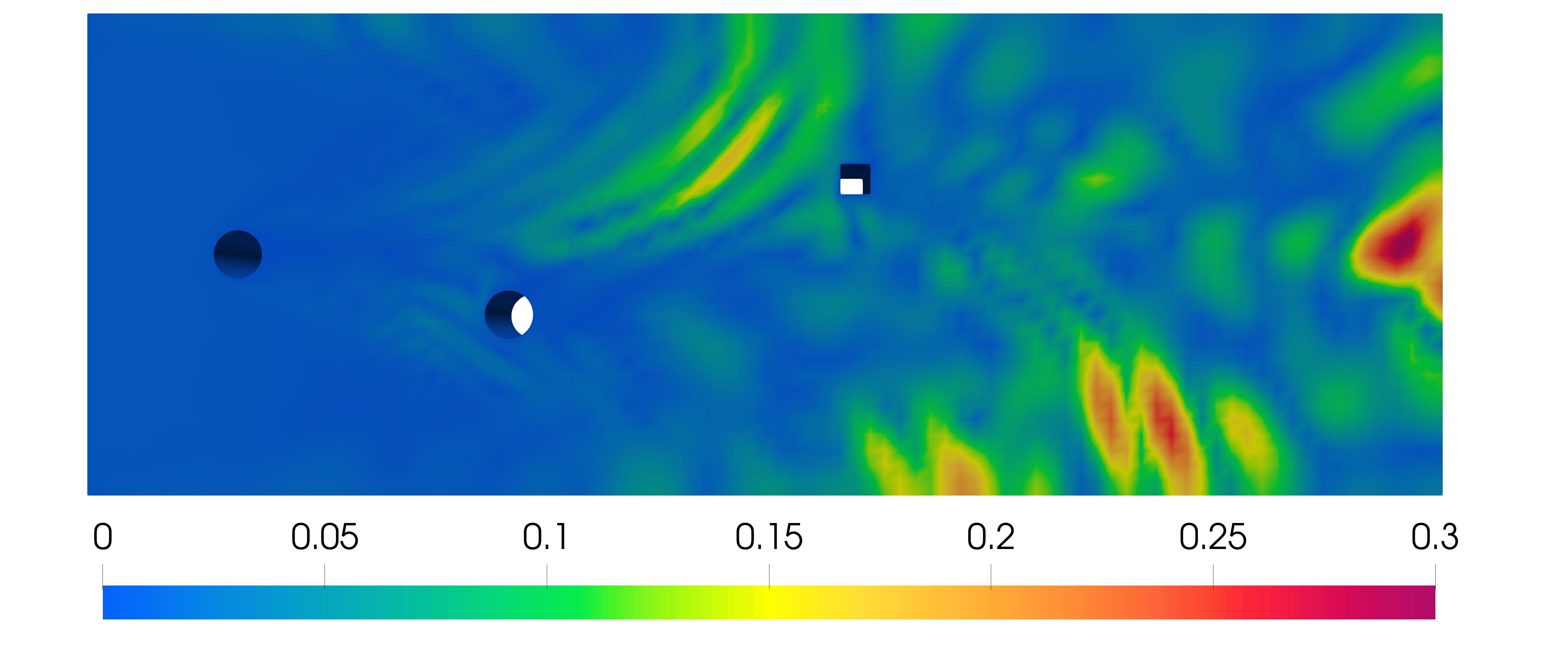}}
	\end{minipage}	
	\begin{minipage}{0.5\textwidth}
		\centering
		\subcaptionbox{t = 90s \label{fig:90error}}
		{\includegraphics[width=6cm]{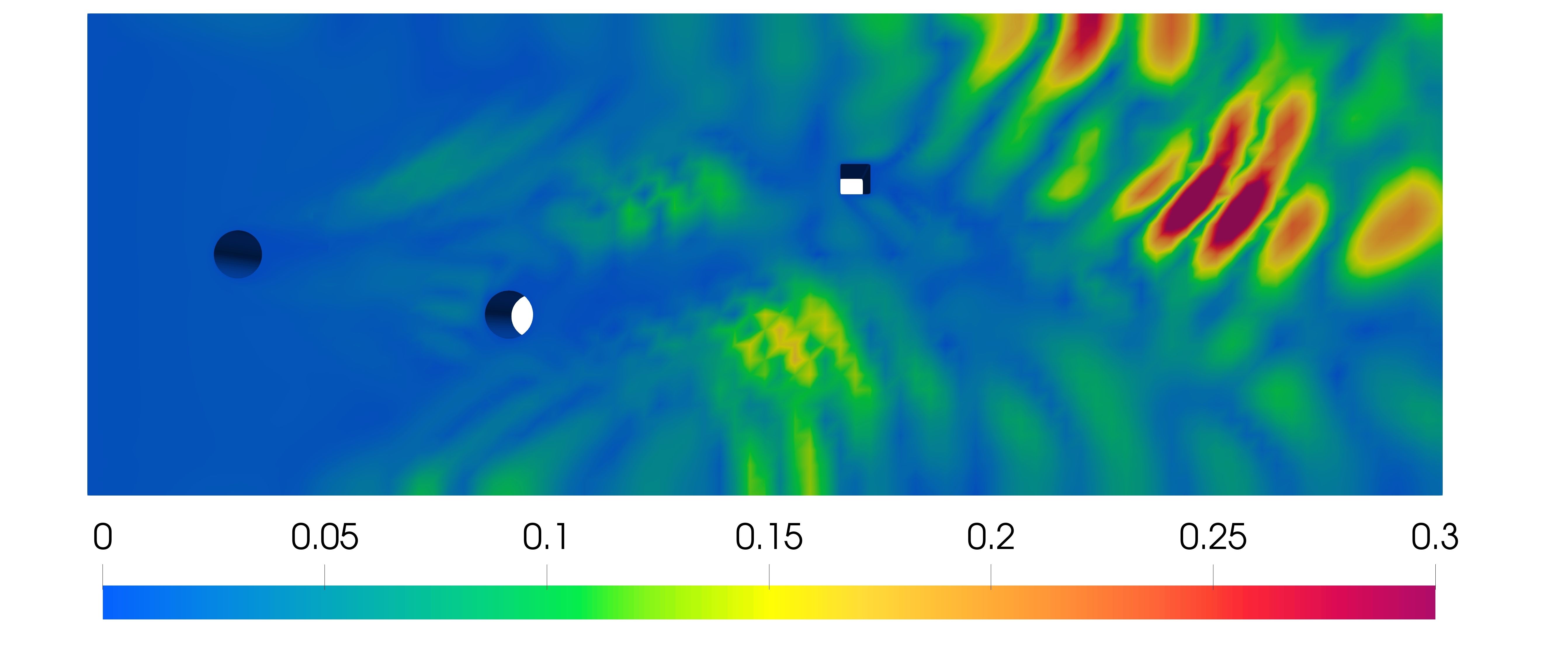}}
	\end{minipage}
	\caption{The velocity difference between solutions of FOM and reconstructions of ROM with 20 POD modes}
	\label{fig:6}
\end{figure}

Comparing Fig. \ref{fig:5} with Fig. \ref{fig:3snapshots}, verified that the model can reconstruct the sinusoidal flow field correctly. The velocity of ROM was subtracted from the velocity of FOM node-by-node to obtain the absolute error value, shown from the top view in Fig. \ref{fig:6}. In most areas, the error of velocity is less than 0.05 m/s. 

It is appropriate to perform the error analysis using the root mean square error (RMSE) and the L2 norm of the relative error, which considers the error at each node and comprehensively reflects the accuracy of ROM.

\begin{equation}\label{eq:20}
	R M S E=\sqrt{\frac{\sum_{i=1}^{k}\left(u_{i}^{FOM}-u_{i}^{ROM}\right)^{2}}{k}},
\end{equation}

\begin{equation}\label{eq:21}
	\frac{\left\|u^{\text {FOM}}-u^{\text {ROM}}\right\|}{\left\|u^{\text {FOM}}\right\|}=\sqrt{\frac{\sum_{i=1}^{k}\left(u_{i}^{FOM}-u_{i}^{ROM}\right)^{2}}{\sum_{i=1}^{k}\left(u_{i}^{FOM}\right)^{2}}},
\end{equation}
where $k$ is the number of mesh nodes. $u_{i}^{FOM}$ is the velocity value of the ith nodes calculated by FOM. $u_{i}^{ROM}$ is the velocity value of the ith nodes calculated by ROM.

\begin{figure}[!htbp]
	\begin{minipage}{0.5\textwidth}
		\centering
		\subcaptionbox{The RMSE\label{fig:RMSE1}}
		{\includegraphics[width=5.8cm]{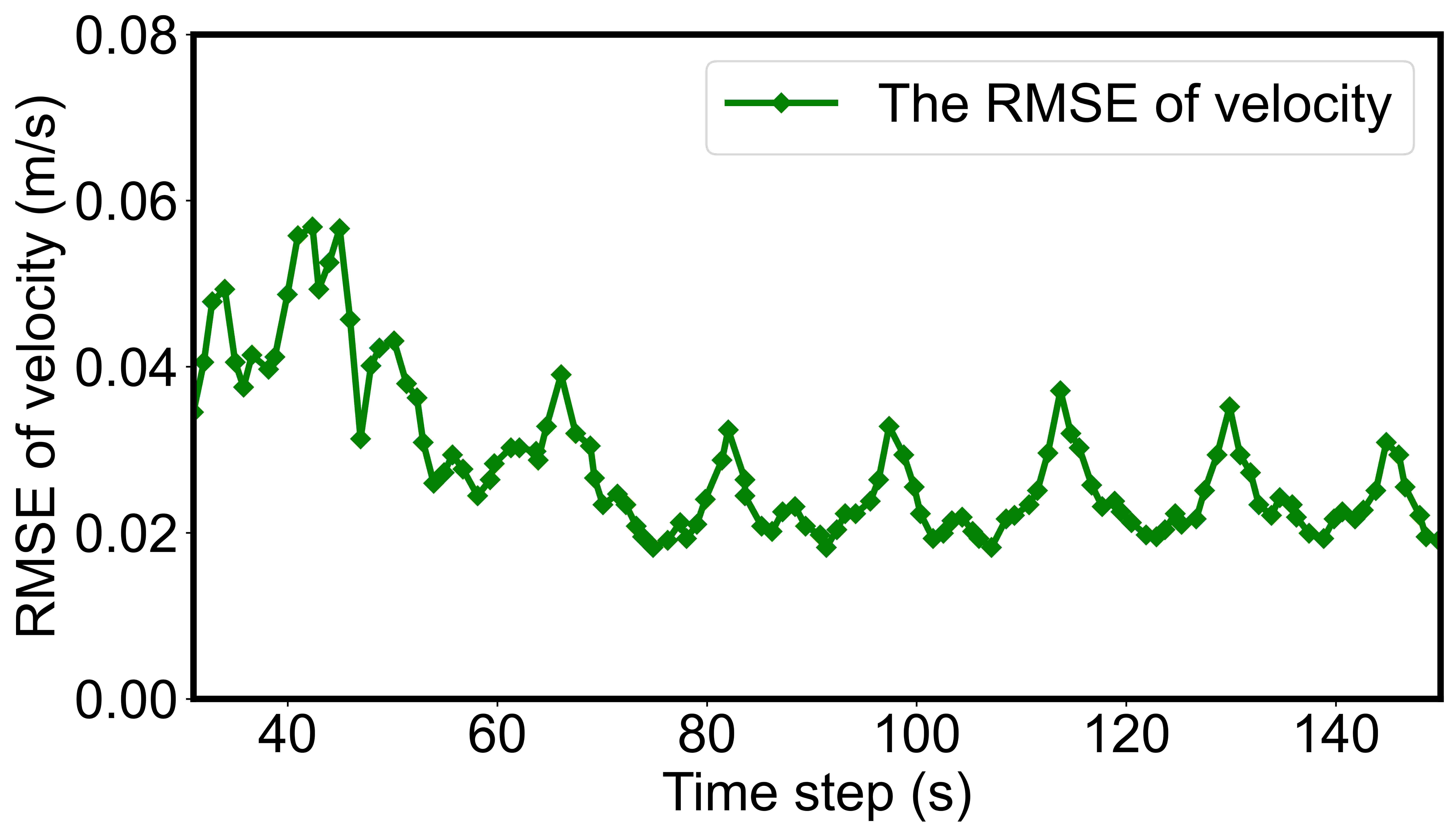}}
	\end{minipage}
	\begin{minipage}{0.5\textwidth}
		\centering
		\subcaptionbox{The L2 norm of the relative error\label{fig:L2E1}}
		{\includegraphics[width=5.8cm]{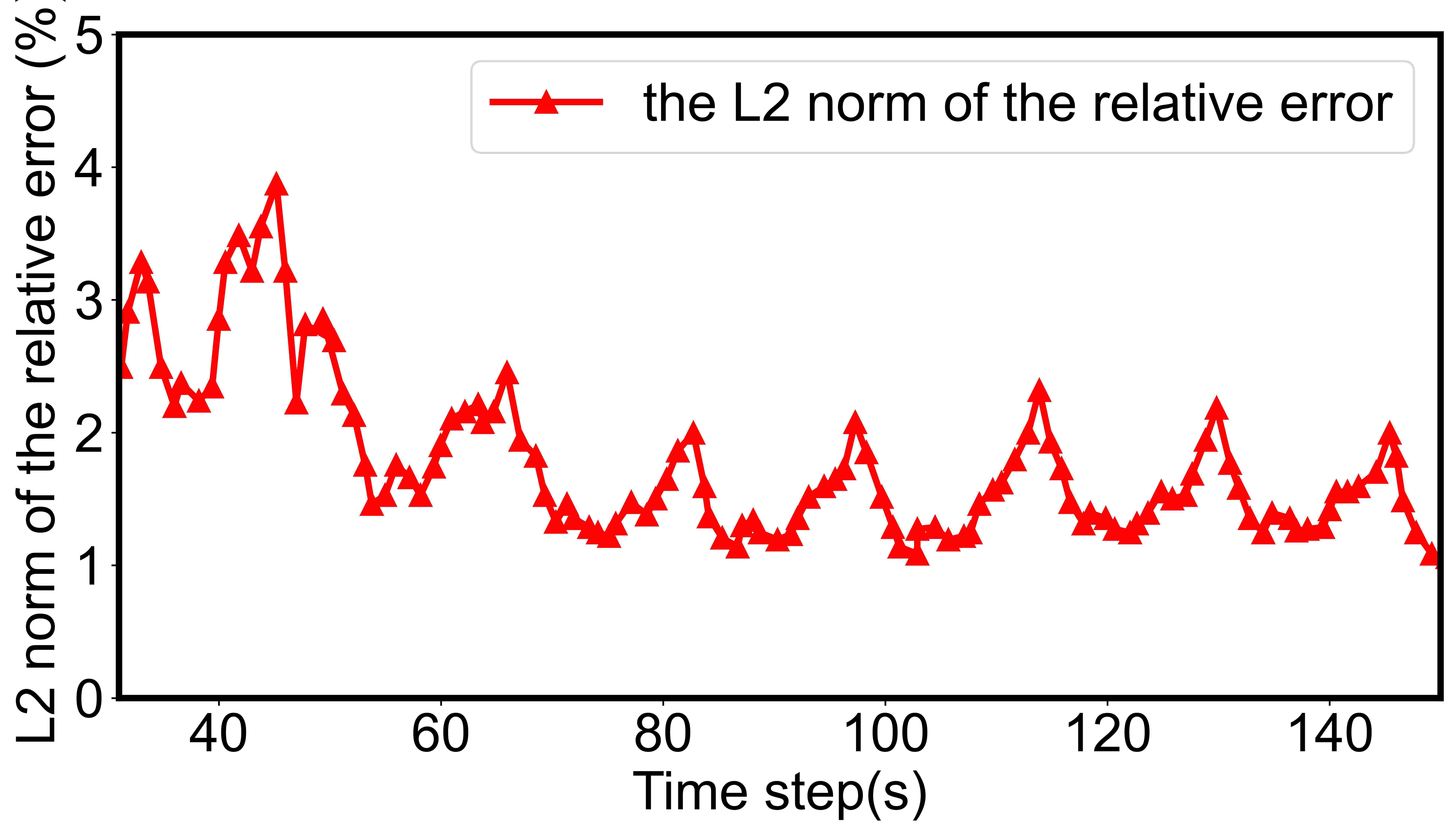}}
	\end{minipage}
	\caption{The RMSE of velocity and The L2 norm of the relative error during (30,150s] (ROM with the first 20 POD modes)}
	\label{fig:7}
\end{figure}

The RMSE of velocity estimated at each time step was plotted as a function of time, as shown in Fig. \ref{fig:RMSE1}. The time variation of the L2 norm of the relative error was determined and plotted as a percentage, as shown in Fig. \ref{fig:L2E1}. The maximum RMSE produced is less than 0.057 m/s, while the maximum L2 norm of the relative errors is less than 3.89\%.

The maximum reconstruction error occurred within a short period when the flow field is not fully affected by the sinusoidal boundary. A higher relative error could be expected during the initial transient.

\begin{figure}[!htbp]
	\centering
	\includegraphics[width=11cm]{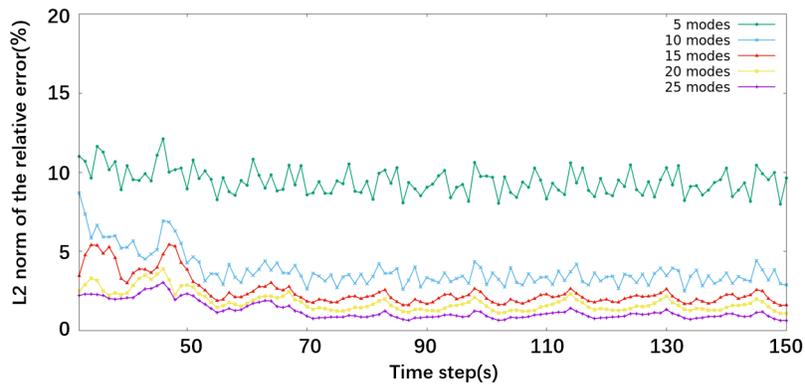}
	\caption{The L2 norm of the relative error for reconstructions by ROMs with 5,10,15,20,25 POD modes}
	\label{fig:pods1}
\end{figure}

To discuss the influence of modes on the accuracy of reconstructions, the error estimation of the ROMs with different modes has been carried out in this work. Fig. \ref{fig:pods1} shows the L2 norm of the relative error of reconstructions over time for ROMs with 5,10,15,20,25 POD modes.

The more modes are retained, the more information of snapshots is retained, and the reconstruction results will be more approximate to the high-precision flow field. 

\subsection{Prediction of velocity fields}

\begin{figure}[!htbp]
	\begin{minipage}{0.5\textwidth}
		\centering
		\subcaptionbox{t = 155s }
		{\includegraphics[width=6cm]{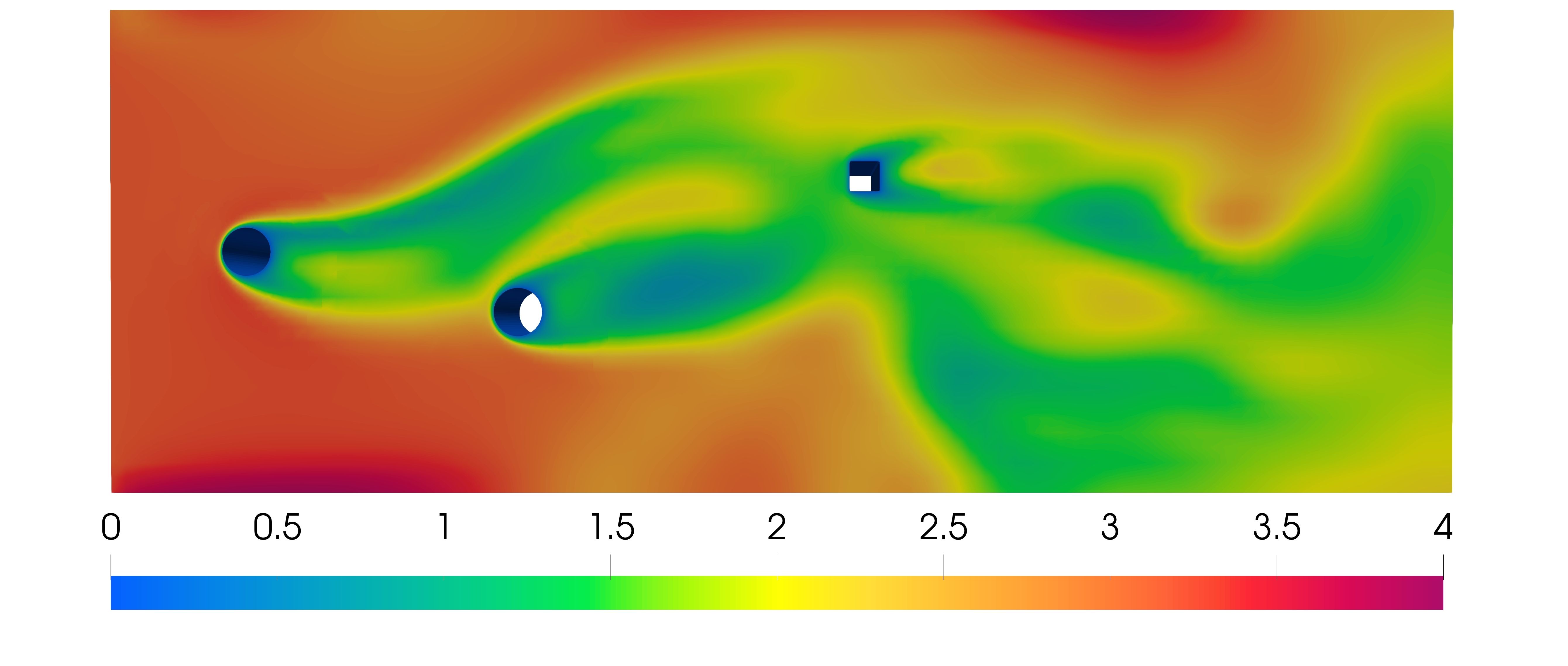}}
	\end{minipage}
	\begin{minipage}{0.5\textwidth}
		\centering
		\subcaptionbox{t = 180s }
		{\includegraphics[width=6cm]{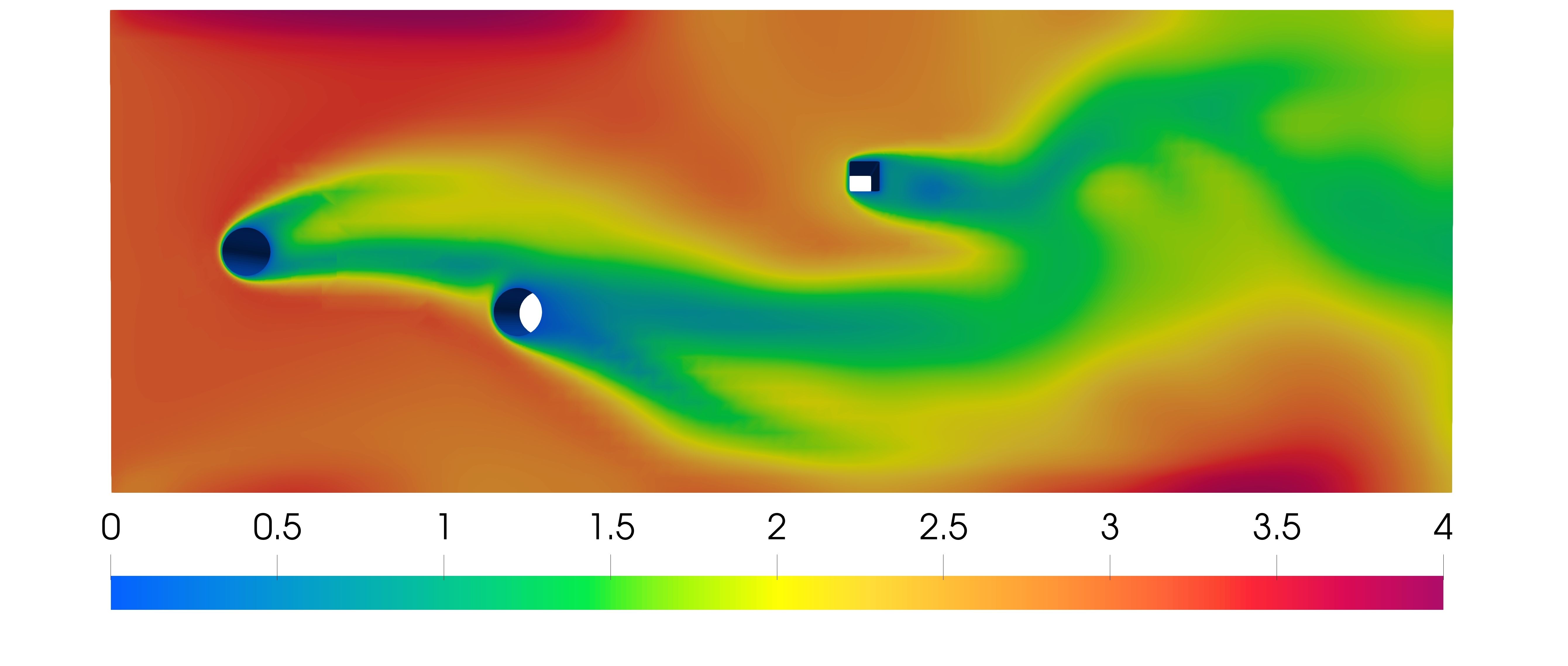}}
	\end{minipage}
	\caption{velocity fields predicted by ROM with 20 POD modes}
	\label{fig:9}
\end{figure}

\begin{figure}[!htbp]
	
	\begin{minipage}{0.5\textwidth}
		\centering
		\subcaptionbox{t = 155s }
		{\includegraphics[width=6cm]{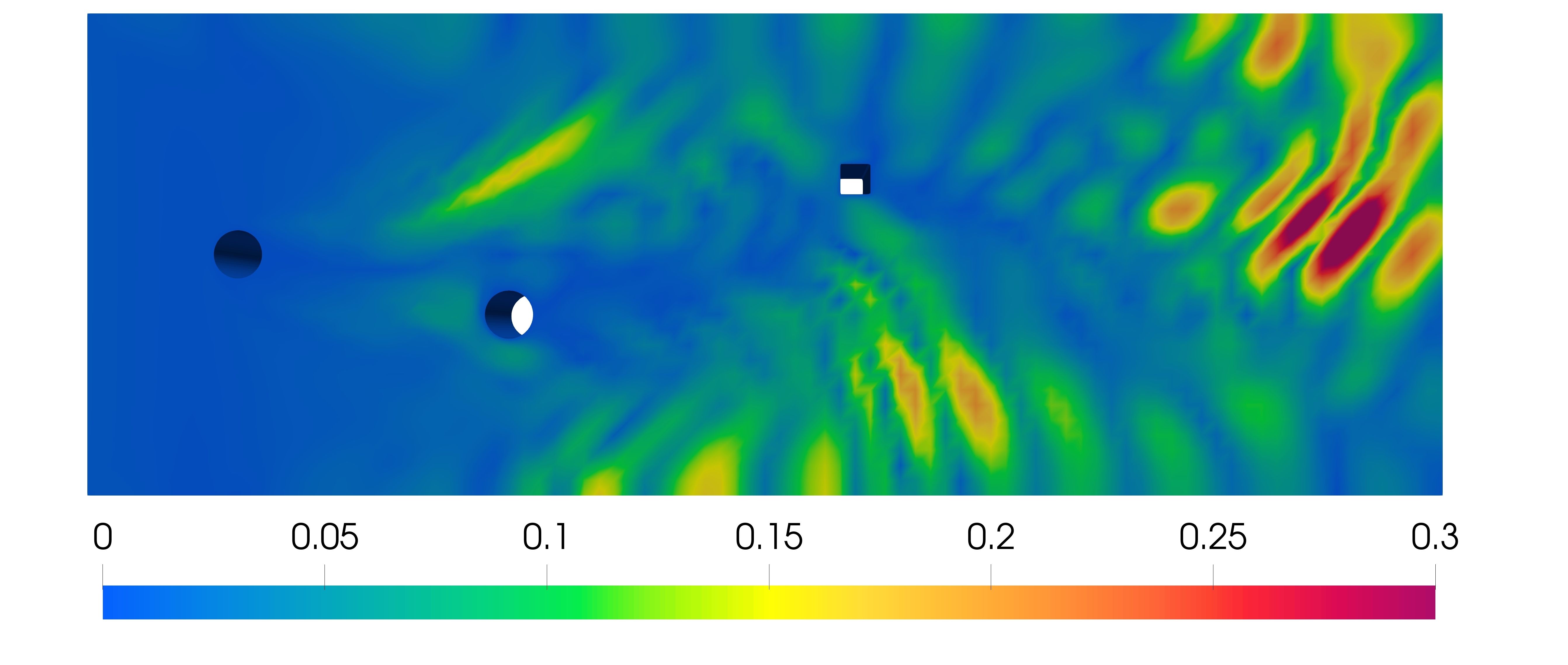}}
	\end{minipage}	
	\begin{minipage}{0.5\textwidth}
		\centering
		\subcaptionbox{t = 180s }
		{\includegraphics[width=6cm]{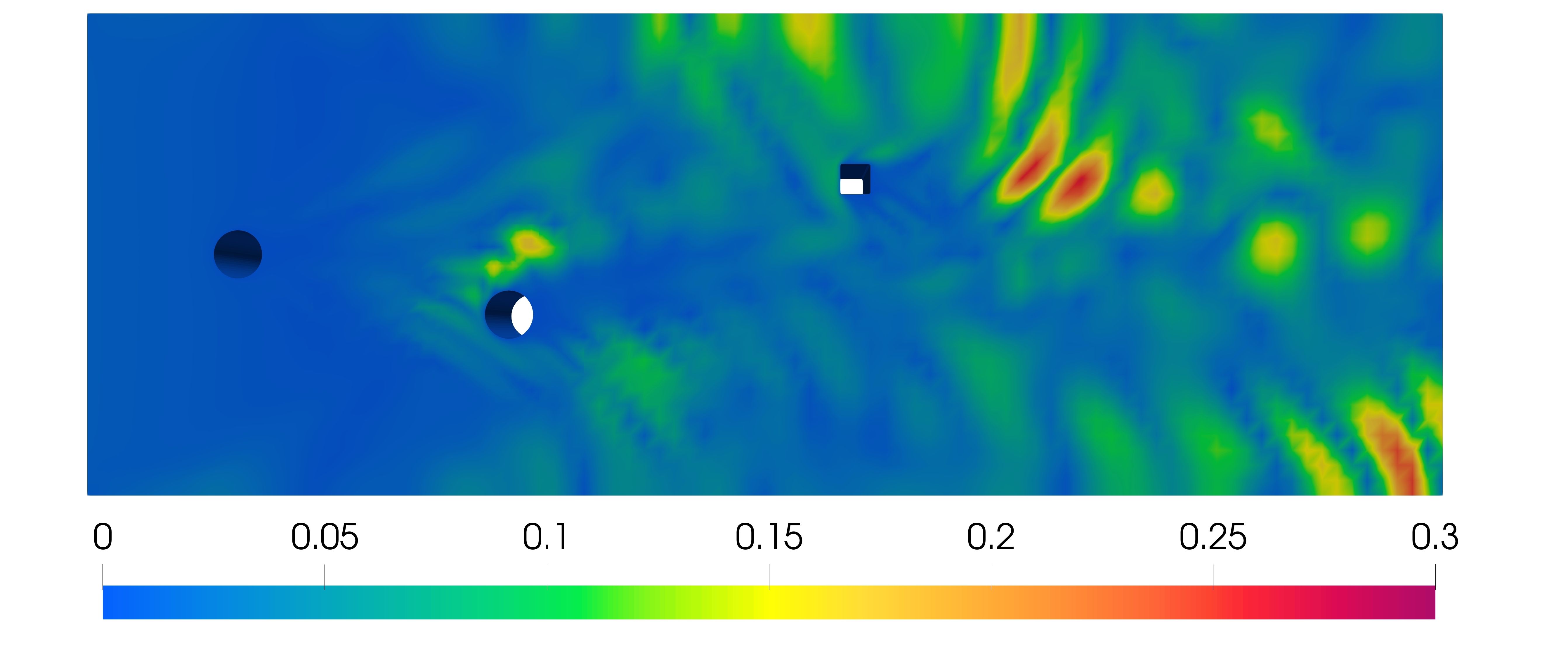}}
	\end{minipage}
	\caption{The velocity difference between solutions of FOM and predictions of ROM with 20 POD modes}
	\label{fig:10}
\end{figure}

In the last subsection, we fitted an interpolation formula to reconstruct the flow field of (30s,150s]. We use the same formula to predict the flow field in the next 80 seconds in the subsection. Fig. \ref{fig:9} shows two prediction solutions of velocity fields at time step t=155s and t=180s calculated by the ROM with 20 POD modes. Then we obtain the high-precision solutions after t=150s by CFD simulation then get the difference between that and predictions of ROM, as shown in Fig. \ref{fig:10}.

\begin{figure}[!htbp]
	\begin{minipage}{0.5\textwidth}
		\centering
		\subcaptionbox{The RMSE\label{fig:RMSE2}}
		{\includegraphics[width=5.8cm]{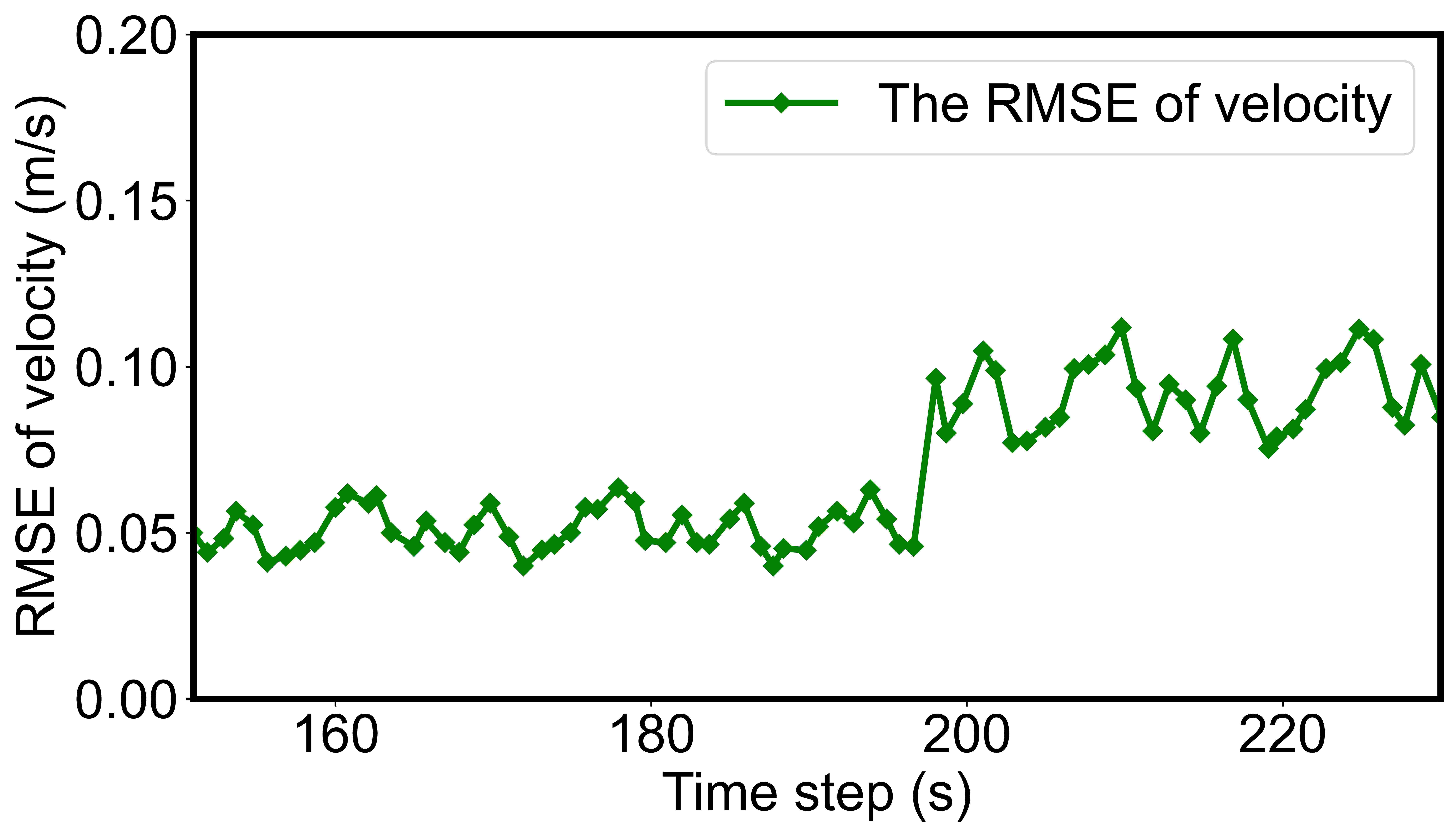}}
	\end{minipage}
	\begin{minipage}{0.5\textwidth}
		\centering
		\subcaptionbox{The L2 norm of the relative error\label{fig:L2E2}}
		{\includegraphics[width=5.8cm]{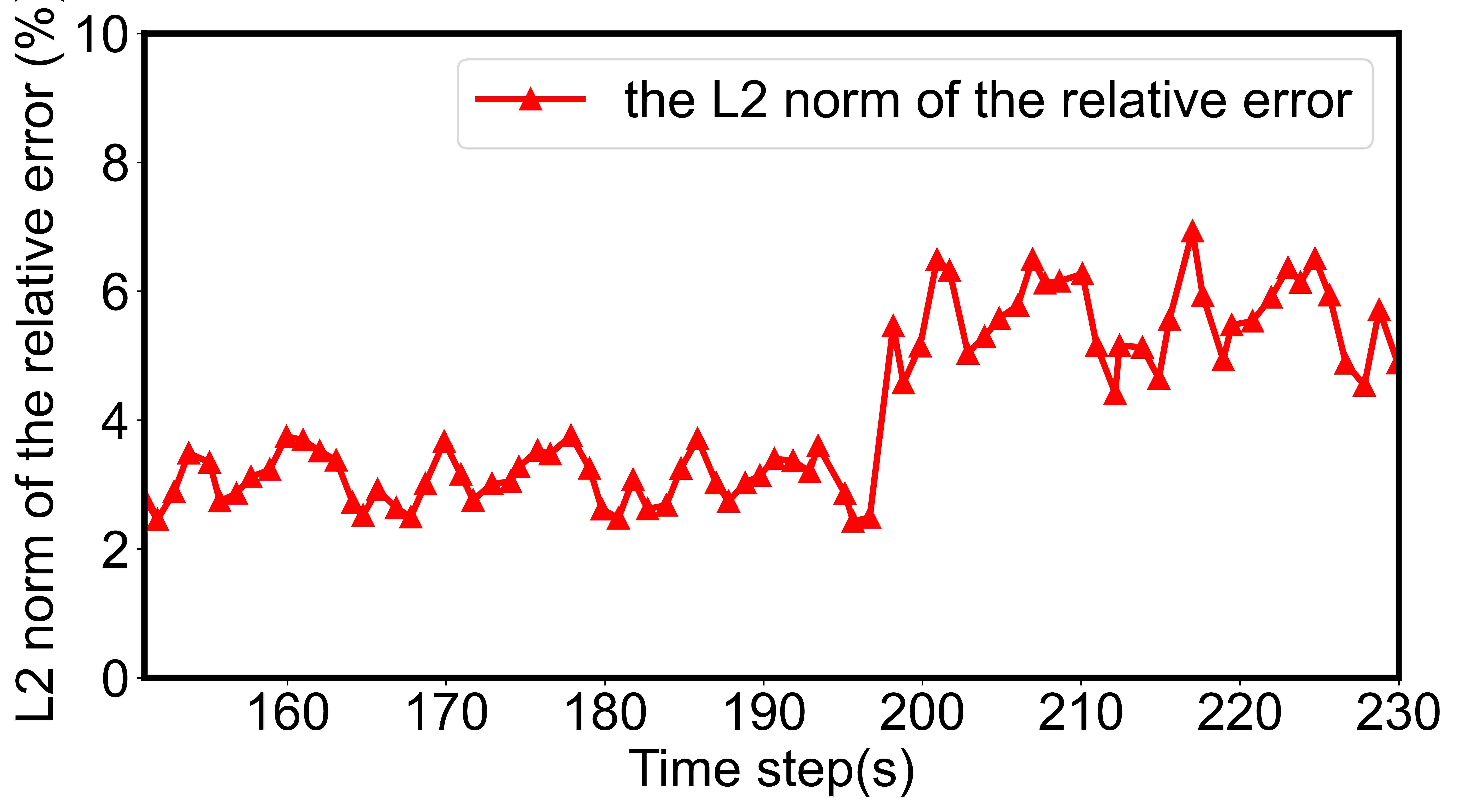}}
	\end{minipage}
	\caption{The RMSE of velocity and The L2 norm of the relative error during (150s,230s] (ROM with the first 20 POD modes)}
	\label{fig:11}
\end{figure}

Comparing the absolute errors in Fig. \ref{fig:6} with Fig. \ref{fig:10}, the error of predictions is much higher than that of reconstructions. Fig. \ref{fig:11} shows the RMSE of velocity and the L2 norm of the relative error versus time in the prediction stage. 

\begin{figure}[!htbp]
	\centering
	\includegraphics[width=11cm]{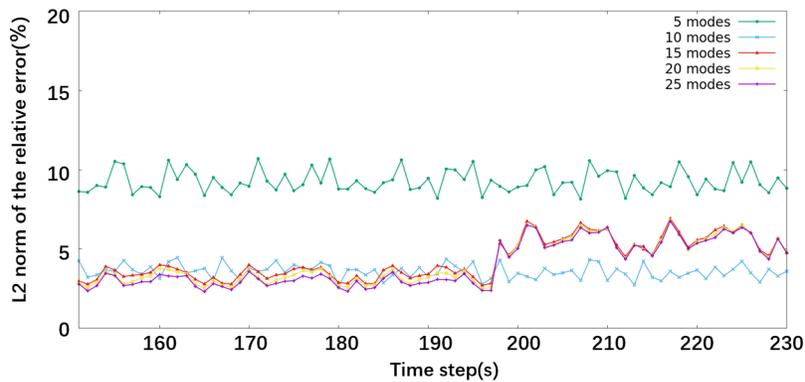}
	\caption{The L2 norm of the relative error for predictions by ROMs with 5,10,15,20,25 POD modes}
	\label{fig:pod2}
\end{figure}

When the ROM was used to predict the solutions, not in the snapshots, the accuracy was significantly decreased. The maximum RMSE is about 0.108 m/s while the maximum L2 norm of the relative errors produced is about 6.82\%. Both errors are twice that of reconstructions and tend to increase suddenly around t=198s. \replaced[id=E]{To}{In order to} find out the cause of the error mutation and the influence of modes on the prediction, we still tested the calculation of different ROMs with different POD modes.Fig. \ref{fig:pod2} shows the L2 norm of the relative error of predictions over time for ROMs with 5,10,15,20,25 POD modes.  

The graph does not conform to the rule that the more ROM modes retained, the better performance is. For ROMs with 5,10 POD modes, the error fluctuates up and down around a constant value. However, for ROMs with 15,20,25 POD modes, the error doubled. \replaced[id=R1]{By comparing the error curve under different number of POD basis functions, we believe this phenomenon can be attributed to overfitting. While increasing the number of POD modes generally improves the accuracy of the reduced-order model during interpolation by capturing more details of the flow field, it can lead to overfitting when extrapolating beyond the training data range. This overfitting causes the model to predict based on noise or specific details of the training set rather than the underlying flow dynamics, resulting in a larger discrepancy between predicted and actual flow fields, as evidenced by the increased L2 norm of the relative error around t=198s. As for the reason why the error increases at a specific time and with a specific number of POD basis functions, it is related to the particularity of this numerical case.}{The possible reason is that too many modes will cause overfitting.}

For the interpolation formula Eq. \eqref{eq:17}, we take the temporal coefficients of POD modes as the training data. The number of modes determines the dimension of the data. When the data dimension is too high, we introduce more noise. The fitted formula Eq. \eqref{eq:19} records many noise characteristics, lowering the prediction accuracy.

\begin{figure}[!htbp]
	\centering
	\includegraphics[width=11cm]{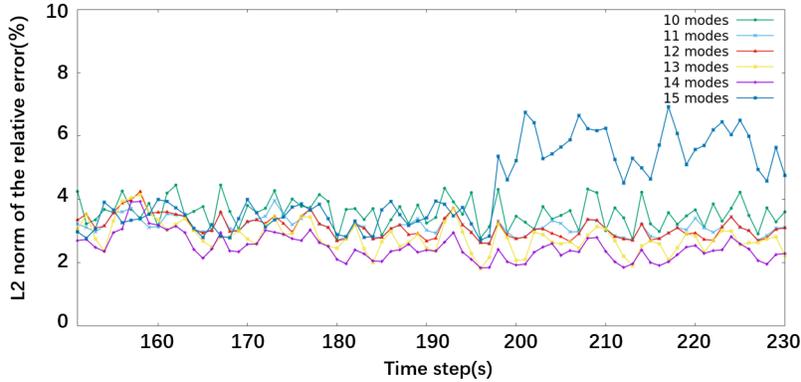}
	\caption{The L2 norm of the relative error for predictions by ROMs with 10,11,12,13,14,15 POD modes}
	\label{fig:pods3}
\end{figure}

It can be seen from Fig. \ref{fig:pod2} that the overfitting occurs between 10 modes and 15 modes. \replaced[id=E]{To}{In order to} find the optimal number of modes, we built ROMs with 11, 12, 13, and 14 modes and obtained their L2 norm of the relative error for predictions, as shown in Fig. \ref{fig:pods3}. 

\replaced[id=R1]{The sudden increase in prediction error occurs at ROM with 15 POD modes. It validates a fundamental principle of ROM: the optimal number of basis functions for extrapolation prediction is always lower than that for interpolation reconstruction. This discrepancy arises from the fundamentally different nature of these two tasks. For flow field reconstruction, increasing mode number monotonically improves accuracy as each additional mode captures finer details of existing data, as shown in Fig. \ref{fig:7}. However, for prediction beyond the training data, excessive basis functions beyond the complexity required to represent underlying physical dynamics will inevitably lead to overfitting. The 15th mode contributes only 0.00012\% of the total velocity field energy, primarily capturing numerical discretization errors and chaotic small-scale wake fluctuations with no predictive value. Adding this mode increases the coefficient vector dimension from 28 to 30, which the 120 training snapshots are insufficient to constrain, resulting in an overly flexible RBF interpolation hypersurface that amplifies noise during extrapolation. The first 14 POD modes thus represent the optimal balance between model expressiveness and generalization ability, capturing 99.91\% of the total velocity field energy and 99.85\% of the total pressure field energy. With this optimal mode number, the maximum RMSE produced is less than 0.071 m/s, while the average RMSE is 0.039 m/s; the maximum L2 norm of the relative errors is 5.13\%, while the average is 2.52\%.}{The overfitting occurs at ROM with 15 POD modes. The first 14 POD modes are optimal for prediction. The maximum RMSE produced is less than 0.071 m/s, while the average RMSE is 0.039m/s. The maximum L2 norm of the relative errors is 5.13\%, while the average is 2.52\%.}

\subsection{Comparison with the POD-Galerkin ROM}

\added[id=R1]{To comprehensively evaluate the performance of the proposed POD-RBF ROM, we conduct a quantitative comparison with the POD-Galerkin method supplied by the open‑source ITHACA‑FV library \cite{ITHACAFV_website}. Its finite-volume-based POD-Galerkin solver has been extensively validated in multiple canonical unsteady flow benchmark cases. Tests are conducted on the built‑in ``04unsteady'' benchmark case of ITHACA‑FV. Offline procedures, including snapshot acquisition, POD basis extraction and modal projection, are kept identical for both ROMs. Only the online flow‑field reconstruction algorithm varies.}

\begin{table}[htbp]
	\centering
	\small
	\setlength{\tabcolsep}{2pt}
	\caption{Comparisons of the POD-RBF model with the POD-Galerkin model.}
	\label{tab:Comparisons}
	\begin{tabular}{ccc}
		\toprule
		ROM & POD-RBF & POD-Galerkin \\
		\midrule
		CPU time& 7.039 s& 12.914 s  \\
		Absolute Error (max)& 3.74e-4 m/s & 5.20e-2 m/s \\
		Absolute Error (mean)&  1.88e-4 m/s & 3.33e-2 m/s  \\
		RMSE (max)& 4.51e-4 m/s & 8.75e-2m/s \\
		RMSE (mean)& 2.91e-4 m/s & 5.84e-2 m/s\\
		L2 relative errors (max)&  0.023\% &  7.47\% \\
		L2 relative errors (mean)&  0.015\%  &  6.26\% \\
		\bottomrule
	\end{tabular}
\end{table}

\added[id=R1]{Table \ref{tab:Comparisons} summarizes the online reconstruction performance metrics. In terms of computational efficiency, the POD-RBF model completes the reconstruction in 7.039 s, representing a 45.4\% reduction in CPU time compared to the POD-Galerkin model (12.914 s). In terms of accuracy, the improvement is even more substantial: the average L₂ relative error of the POD-RBF model is 0.015\%, which is 417 times lower than the 6.26\% achieved by the POD-Galerkin model. All error metrics show improvements of more than two orders of magnitude.}

\added[id=R1]{The observed performance differences arise from the fundamental methodological distinctions between the two approaches. The POD-Galerkin method is an intrusive ROM that requires projecting the Navier-Stokes equations onto the POD subspace and solving a system of coupled reduced-order ordinary differential equations (ODEs). This process introduces numerical dissipation and truncation errors, particularly for unsteady flows with complex vortex dynamics. In contrast, the POD-RBF model is a non-intrusive data-driven framework that replaces the Galerkin projection with radial basis function interpolation. This eliminates the need for equation manipulation or reduced-order ODE solution, avoiding projection-induced numerical errors while simplifying implementation and reducing computational cost.}

\section{Conclusion}

This paper presented a reduced-order model based on POD and RBF to reconstruct and predict unsteady flow with periodically varying boundary conditions. The numerical case is a three-dimensional flow around cylinders. In particular, the velocity of inlet flow varies sinusoidally.  

We use the simulation during (30s,150s] to obtain the snapshots for the POD modes generation because this time is long enough to include the periodic change of inlet velocity completely. Within this time window, the snapshots are collected every 1 s, returning a total number of 120 snapshots. We reserved the first 20 POD modes that constitute a major part of system energy to create the ROM based on RBF interpolation. Applying the ROM to calculate solutions during the time period of (30, 230s], we obtain 120 reconstructions and 80 predictions.

The velocity fields calculated by ROM are consistent with the full order model. The effect of prediction is worse than that of reconstruction. For reconstruction, the accuracy of the ROM becomes higher when the number of modes increases. However, for prediction, building the ROM with too many modes will lead to an overfitting problem, reducing the accuracy. In this case, the optimal number of modes for prediction is 14, and the average L2 norm of the relative errors is about 2.52\%.

The ROM calculations are performed on a laptop computer, with an Intel Core i5-8300H running at 2.3GHz. While the FOM calculations uses a cluster of 64 processors, with Intel Xeon Gold 6130 running at 2.1GHz. The total CPU time for FOM simulation of the flow field in 200s is about 20034.5s. In comparison, the CPU time for ROM is about 62.07s, which includes generating the modes and calculating the coefficients.

The ROM built by proper modes can predict solutions for the periodical flow field with a periodical boundary condition accurately.  The advantage of this method is stability and it only relies on the snapshots, which can be obtained from both the numerical simulations and experimental datas. It is demonstrated that accuracy is maintained while CPU time is reduced by more than 99\% in comparison with the FOM. 

\added[id=R1]{Furthermore, comparative tests against the native POD-Galerkin model from the open-source ITHACA-FV library are performed. With identical offline modeling settings, the proposed POD-RBF non-intrusive ROM exhibits superior comprehensive performance, including higher reconstruction accuracy and lower online computational cost, compared with the classic intrusive POD-Galerkin method.}

\added[id=R2]{Notably, the proposed POD-RBF reduced-order model has clear applicable scenarios and generalization boundaries. The model exhibits excellent interpolation reconstruction performance for both periodic and non-periodic boundary conditions within the training parameter space. However, reliable long-term extrapolation prediction of the model is only constrained to unsteady flow problems with fixed or periodically varying boundary conditions. Due to the inherent characteristics of RBF interpolation and the periodic feature of the POD snapshot training dataset adopted in this work, the model cannot achieve accurate extrapolation prediction for flow fields under non-periodic boundary conditions. This limitation also points out the future optimization direction of this method, that is, expanding the model generalization ability to adapt to complex aperiodic flow boundary conditions.}

\section*{Acknowledgements}

This work was partially supported by the national key research \& development (R\&D) plan under Grant No.2017YFC0803300 and the National Natural Science Foundation of China (Grant No: U2033206).

\bibliography{mybibfile}

\end{document}